\documentclass[aps,floatfix,groupedaddress,nofootinbib,superscriptaddress,twocolumn]{aastex701}

\usepackage{tabularx}              
\usepackage{array}                
\usepackage{amsmath}
\usepackage{bm}
\usepackage{graphicx}
\usepackage{booktabs}
\usepackage{xcolor}

\newcommand{\cl}{C_\ell}
\newcommand{\rij}{r_{ij}}

\newcommand{\ha}{\ensuremath{\mathrm{H}\alpha}}
\newcommand{\hb}{\ensuremath{\mathrm{H}\beta}}
\newcommand{\oii}{\ensuremath{[\mathrm{O}\,\text{\textsc{ii}}]}}
\newcommand{\oiii}{\ensuremath{[\mathrm{O}\,\text{\textsc{iii}}]}}
\newcommand{\lya}{Ly$\alpha$}

\shortauthors{Roy et al.}

\begin{document}

\title{Reading Between the Lines: Forward Modeling Dust, Continuum, and Spectral Cleaning for Multi-Line Intensity Mapping}

\shorttitle{Between the Lines: Multi-Line Intensity Mapping}

\author[0000-0001-5729-0246]{Anirban Roy}
\affiliation{Department of Physics, New York University, 726 Broadway, New York, NY, 10003, USA}  
\affiliation{Center for Computational Astrophysics, Flatiron Institute, New York, NY 10010, USA}
\email{ar8816@nyu.edu}

\author[0000-0002-6748-6821]{Rachel S. Somerville}
\affiliation{Center for Computational Astrophysics, Flatiron Institute, New York, NY 10010, USA}
\email{rsomerville@flatironinstitute.org}

\author[0000-0002-2091-8738]{Anthony Pullen}
\affiliation{Department of Physics, New York University, 726 Broadway, New York, NY, 10003, USA}  
\affiliation{Center for Computational Astrophysics, Flatiron Institute, New York, NY 10010, USA}
\email{anthony.pullen@nyu.edu}

\begin{abstract}
Line intensity mapping (LIM) offers a tomographic view of galaxy evolution by
measuring the aggregate emission from unresolved galaxies. In the optical and near-infrared, the line emission is accompanied by much brighter, spectrally smooth continuum emission that must be removed to recover the line auto- and cross-power spectra. We develop a forward-modeling framework to
study this problem using a $\sim 2\,\mathrm{deg}^2$ lightcone drawn from a
cosmological $N$-body simulation populated with galaxies using a physics-based
semi-analytic model (SAM) of galaxy formation. We construct intensity maps for
the stellar continuum and for the strongest optical lines, \ha, \hb,
\oiii\ $\lambda5007$, and \oii\ $\lambda3727$, including nebular dust
attenuation tied to galaxy properties. From these spectral cubes, we measure
cross-channel angular power spectra, $\cl(\lambda_i,\lambda_j)$, and the
normalized correlation matrix $\rij$. In line-only maps, the correlation
matrices show the expected same-redshift ridges between different emission
lines, demonstrating how multi-line intensity mapping (MLIM) can isolate large-scale structure and
probe galaxy properties such as dust attenuation. We show that dust suppresses
the line cross-power by an amount that depends on galaxy properties, wavelength,
and line pair, so a single overall amplitude cannot capture its effect. In
total maps, however, the continuum dominates the raw correlations and hides
much of the line-ridge structure. We therefore apply principal component
analysis (PCA)-based spectral cleaning and quantify the resulting line-transfer
function using the simulation truth. Removing approximately 20 PCA modes gives
the best trade-off between line recovery and continuum suppression in our mock
maps. Our results demonstrate both the promise and the challenges of extracting
dust-sensitive LIM observables from SPHEREx-like observations, and highlight
the need to model continuum cleaning and its transfer function as part of any
quantitative inference pipeline.
\end{abstract}

\keywords{Line intensity mapping (2084), Large-scale structure of the universe (902), Interstellar dust extinction (837), Galaxy evolution (594), Near infrared astronomy (1093), Astronomical simulations (1857)}

\section{Introduction}

The interstellar medium of galaxies produces a rich array of emission lines, created when atoms and ions are collisionally or radiatively excited and then cascade to lower energy states.
The cosmic histories of star formation, metal enrichment, as well as the progression of reionization, are encoded in these emission lines 
\citep{Khostovan2015, Gong2017, Fonseca2017}. Much of this emission
arises from faint sources that lie below the detection thresholds of
conventional galaxy surveys, yet in aggregate these galaxies make a
major contribution to the budgets of ionizing photons, star formation,
and metal production \citep{Madau2014,Robertson2015}. Line intensity
mapping (LIM) sidesteps the need to detect individual galaxies by
measuring spatial fluctuations in the integrated line intensity field,
which trace the summed emission from both resolved and unresolved
sources \citep{Kovetz2017LIM_report,Kovetz2019,Bernal2022}. Over the past decade,
LIM has emerged as a promising route to mapping large-scale structure at
redshifts that are challenging for resolved spectroscopic surveys, with
active programs targeting the 21\,cm line at radio frequencies,
[C\,\textsc{ii}] and CO in the millimeter and submillimeter, and rest-frame optical lines at near-infrared wavelengths \citep{Suginohara1998, Righi2008b, Lidz2011_CO, Carilli2011, Fonseca2017, Gong2017, Kovetz2017LIM_report, Chung2018CII, PadmanabhanCO, Padmanabhan_CII, Dumitru2018, Chung2018CO, Kannan:2021, Murmu:2021ljb, Karoumpis2021, Limfast2, slick-garcia}.

The rest-frame optical and near-infrared window is particularly promising for LIM, since it contains multiple strong nebular emission lines, namely \ha, \hb, \oii $\lambda3727$, and
\oiii\ $\lambda5007$  \citep{Silva2018, Limfast1}. The Balmer lines (e.g. \ha\ and \hb) are recombination lines whose
luminosities track the production rate of hydrogen-ionizing photons, and
therefore the recent (past $\sim 10$--20 Myr) star-formation rate \citep{Kennicutt1998},
whereas the strong forbidden lines of oxygen are collisionally excited
transitions whose luminosities are sensitive to gas phase
metallicity and ionization state \citep{Kewley2013}.
A spectroscopic survey covering
the observed-frame optical/Near-IR ($\sim0.75$--$5~\mu{\rm m}$), of the kind enabled by all-sky spectral
mapping with SPHEREx \citep{Dore2014, Dore2018}, covers these lines over complementary redshift intervals spanning much of $z\sim0$--$5$, over which the bulk of cosmic star
formation took place. Because each line enters the observed band from a different redshift, a single survey delivers several distinct line tracers as a function of cosmic time, and any two lines originating in the same comoving volume subtended by the instrument can be cross-correlated \citep{Schaan:2021gzb}. Such cross-correlations isolate the clustering signal common to both maps while suppressing interloper lines and instrumental systematics that afflict any individual line \citep{Roy-Lim-LLX}.

The conceptual centerpiece of MLIM is the cross-channel power
spectrum $\cl(\lambda_i,\lambda_j)$, the angular power spectrum between
the intensity fluctuations observed in wavelength channels $\lambda_i$
and $\lambda_j$ \citep{cheng2024}. This matrix encodes both the auto-spectra, recovered on
the diagonal $\lambda_i=\lambda_j$, and the full set of cross-spectra
between distinct channels, and it is the latter that carry the
multidimensional information specific to an intensity mapping survey. Because a
single observed channel collects emission from several rest-frame lines
originating at different redshifts, no individual channel corresponds to
a clean redshift slice of the underlying field. The two-dimensional
matrix of channel-pair spectra, however, carries a distinctive geometric
signature as two channels are correlated whenever they are sourced by a
pair of different rest-frame lines emitted at a common redshift. For two
lines $a$ and $b$, this same-redshift ridge satisfies
\begin{equation}
  \lambda_b = \lambda_a\,\frac{\lambda_{b,{\rm rest}}}{\lambda_{a,{\rm rest}}},
  \label{eq:ridge}
\end{equation}
where both observed wavelengths lie within the instrumental spectral range.
Each ridge is therefore a geometrically labeled probe of a specific line
pair at a specific redshift, and the slope of a ridge fixes the ratio of
the rest wavelengths of the two lines. This structure
underlies proposals to use multi-line cross-correlations for line
separation, redshift tomography, and constraints on the physics of emission lines
\citep[e.g.,][]{Visbal2010,Gong2014,Gong2017,Cheng2016,Cheng2020}. The
same geometry also aids foreground control, since continuum foregrounds
and interloper lines have a smooth or otherwise distinct frequency
response that does not reproduce the sharp same-redshift ridges of the
target line pairs \citep{Switzer2019}.

Several obstacles separate this promising concept from a working
inference pipeline. These have typically been studied in isolation, and taking a first step towards
treating them together is the focus of this work. First, dust
attenuation reshapes the line emission in a way that depends strongly on
wavelength, galaxy properties, and geometry \citep{SalimNarayanan2020}. 
The attenuation experienced by emission from nebular gas differs from that of the stellar continuum \citep{Calzetti2000, CharlotFall2000, Reddy2020}, and the attenuation curve is not universal \citep{Salim2018,SalimNarayanan2020,Sommovigo2025}. This affects the ratios of line strengths in a complex manner that depends on galaxy properties and cannot be captured via a single correction factor, though many forecasts either treat dust implicitly or through such a fixed mean
correction factor. Second, optical and near-infrared LIM is not free of
foregrounds. The unresolved stellar continuum is bright, spectrally smooth, and strongly coherent across channels, so that in the absence of continuum removal the cross-channel matrix is dominated by this smooth component and the line
ridges are obscured \citep{vancuyck2023, Cagliari2025}. Continuum
cleaning is therefore essential to recover the LIM signal. Third, this
cleaning is in general not signal preserving. Any filtering aggressive enough to
remove the smooth continuum also attenuates a fraction of the line
fluctuations, so the recovered cross-spectra underestimate the true power unless the associated transfer function is calibrated and propagated into the
inference rather than assumed to be unity \citep{Cunnington2023}. In this work, we bring all of these effects together in a single forward model and quantify how they jointly shape the cross-channel matrix.

In this work, we present an end-to-end forward model of optical/NIR MLIM statistics, in which the line emission, dust attenuation, continuum emission, map making, and spectral cleaning are all considered within a physically motivated framework. We begin with a lightcone generated by a cosmological $N$-body simulation and populate it
with galaxies using a physics-based semi-analytic model. The known physical properties of each galaxy are used to assign its nebular line luminosities, which are then attenuated using a dust model driven by those same properties. Both the line and continuum emission are projected into observed-frame spectral maps. This model avoids representing the LIM signal
as merely a set of line amplitudes, tracing instead the galaxy-level physics that produces these statistics in a realistic setting of cosmological structure formation.

The main objective of this work is to analyze how the multi-line signal is
modified by several effects that are frequently studied individually: dust,
which changes the relative intensities of the various lines; continuum
emission, which acts as a bright, coherent foreground; spectral cleaning,
which modifies the cosmological signal itself; and finite survey geometry,
which alters the measured angular correlations. Using mock maps and a common
analysis pipeline, we track each of these effects from the intrinsic line
fields through to the continuum-dominated maps and finally to the cleaned
cross-channel correlation matrix. This lets us determine not only whether the
same-redshift line ridges are present, but also how robustly they endure
through the sequence of observational operations required to construct a
realistic map. By establishing the link between galaxy-formation parameters
and the observed spectral statistics, our approach provides a basis to
investigate which lines carry information on star-formation rates and dust
properties, which cleaning operations preserve or erase the signal, and which
transfer functions must be carried forward into the interpretation of
SPHEREx-like surveys.

The general goal of this work is to address whether MLIM can probe
the physical properties of the unresolved galaxy population rather than just
aggregate line power. In a spectral data cube, different rest-frame lines
sample the same large-scale structure at different observed wavelengths.
Their cross-channel correlations therefore encode not only the total line intensity, but also the relative
strengths of the lines, their redshift evolution, and the effects of dust
attenuation.  If continuum can be removed without erasing the
line signal, the wavelength--wavelength LIM matrix becomes a compressed
observable for galaxy physics as it can test line-luminosity calibrations,
constrain effective nebular attenuation, and connect the integrated
emission field to star formation in galaxies too faint to detect
individually.

The rest of the paper is organized as follows.
Section~\ref{sec:sam_sims} introduces the simulated galaxy catalog used in
this work and describes how we convert the galaxy population into
observed-frame intensity and continuum maps. We summarize the construction of the wavelength-channel grid for an example experiment,
the emission-line luminosity model, the treatment of nebular dust
attenuation, and the construction of the continuum maps.
Section~\ref{sec:experiments} describes the experimental setup and defines the cross-channel angular power spectra and
normalized correlation matrices that form the main statistics of our
analysis, and describes the estimator pipeline used to measure them from
the maps. Section~\ref{sec:results} presents the results of the
multi-line correlation signal before continuum cleaning, including the
location and strength of the same-redshift line ridges, the effect of
dust attenuation, and the contribution of shot noise. Section~\ref{sec:cleaning} then introduces the
continuum-cleaning methods considered in this work and quantifies their
impact on both continuum suppression and line-signal recovery. Section~\ref{sec:inference} focuses on parameter inference from both the
raw and the principal component analysis (PCA) based cleaned data: the raw maps retain the continuum and
therefore constrain its residual amplitude and tilt, while the cleaned
maps suppress the continuum and isolate the line physics.
Section~\ref{sec:discussion} discusses the implications of these results
for interpreting optical/NIR MLIM measurements, and
Section~\ref{sec:conclusion} summarizes our main conclusions. Throughout this work, we assume a flat $\Lambda$CDM cosmology with
parameters fixed to values consistent with the Planck TT,TE,EE+lowE+lensing analysis
\citep{P18:main}.

\section{Modeling of Galaxy Properties: Stellar Continuum, Line emission, and Dust}\label{sec:sam_sims} 

\subsection{Semi-analytic Lightcone Catalog}
\label{sec:sam_catalog}

We base the mock observations on a 2 deg$^2$ mock lightcone galaxy catalog presented in \citet{Yung2023} and publicly available at \url{https://flathub.flatironinstitute.org/group/sam-forecasts}. Dark matter halos were extracted along a past lightcone from the Small Multidark Planck (SMDPL) cosmological N-body simulation from the Multidark suite \citep{Klypin2016}. Each halo was populated with galaxies using the Santa Cruz semi-analytic model (SC SAM), which constructs a semi-analytic halo merger history and applies physically
motivated prescriptions to evolve the baryonic components of galaxies,
including gas cooling, star formation and stellar feedback, chemical enrichment, and black-hole growth and feedback  \citep{Somerville2008, Somerville2015, Yung2023}. The Santa Cruz SAM therefore provides, for each
galaxy, a self-consistent and connected set of physical and observable properties, including
stellar mass, star-formation rate, gas content, metallicity, dust
attenuation parameters, and synthetic photometry. The photometry is
computed by applying stellar population synthesis and dust attenuation
models to the predicted star-formation and enrichment histories \citep{Somerville2008, Yung2023}. The physical processes in the Santa Cruz SAM have been calibrated to reproduce key observables at $z\sim 0$, and the predictions of the model have been extensively tested and shown to reproduce observations from galaxy surveys over a wide range of redshift and galaxy mass \citep{Somerville2008, Somerville2015, Yung2019, Yung2023}.

\begin{figure*}[htbp]
    \centering
    \includegraphics[width=0.95\linewidth]{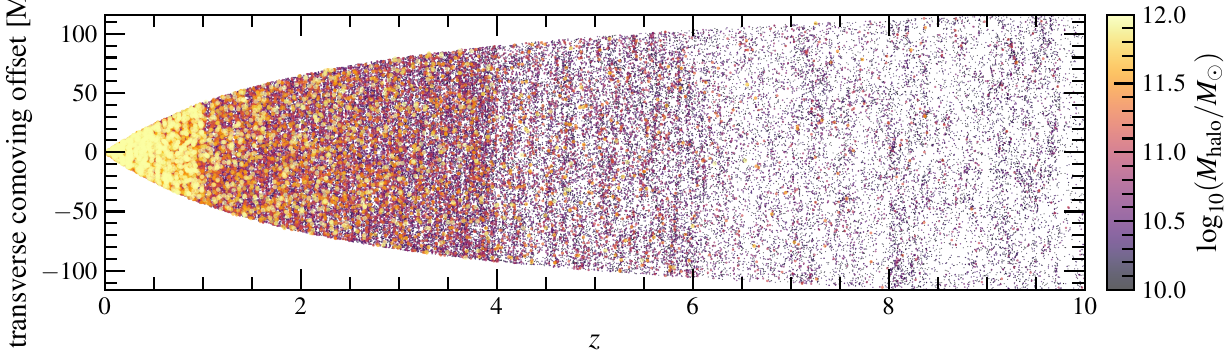}
    \caption{
Geometry of the SAM lightcone catalogue. Each point is a halo, plotted
by redshift and transverse comoving offset from the field center and
colored by mass. For
clarity, only a random subsample of the full catalogue is plotted. The widening
of the cone with redshift reflects the fixed angular footprint projected
into comoving coordinates, while the concentration of massive haloes at
low redshift reflects hierarchical growth of structure.
}
\label{fig:sam_lightcone}
\end{figure*}

\begin{figure}[htbp]
  \centering
  \includegraphics[width=\columnwidth]{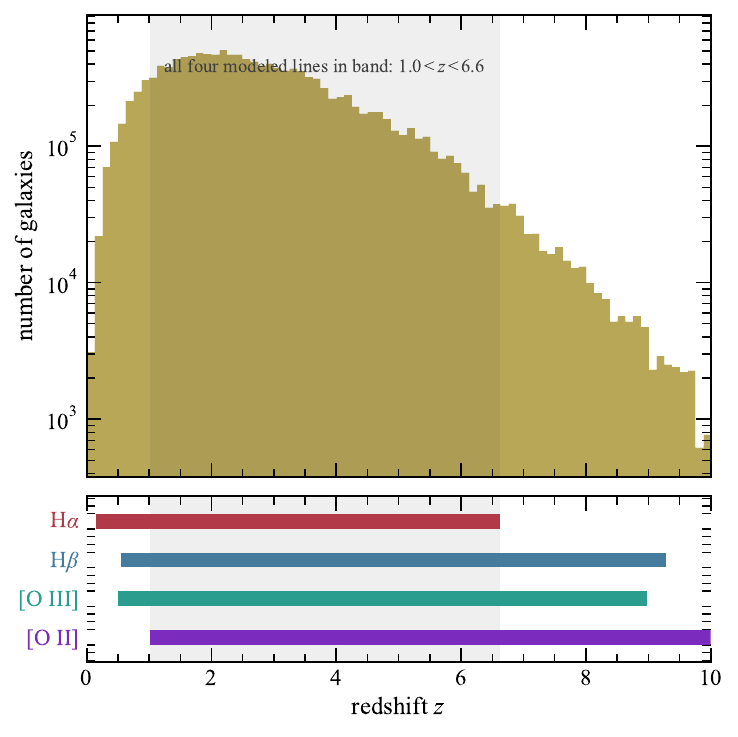}
  \caption{Redshift coverage of the SAM lightcone and the modeled rest-frame optical lines. The histogram shows the galaxy redshift distribution in the catalog, while the colored bars mark the redshift intervals over which \ha, \hb, \oiii, and \oii\ fall in the $0.75$--$5~\mu{\rm m}$ observed-frame band. The shaded region marks the interval $1.0<z<6.6$ where all four modeled lines are simultaneously observable, setting the redshift range in which same-redshift multi-line ridges can appear in the channel-correlation matrices.}
  \label{fig:samprops}
\end{figure}

This catalog is well suited to our purpose because the same galaxy
population can be used to generate the line emission, continuum emission,
dust attenuation, and clustering signal. Many LIM forecasts assign line
luminosities statistically, often through average empirical relations between halo
mass, star-formation rate, and line luminosity \citep{Roy2023-limpy}. Here we instead use the
SAM catalogs to enable a controlled galaxy-level analysis in the lightcone set-up: each object has a sky
position, redshift, and physical properties from which its contribution
to the observed-frame intensity maps can be constructed. The embedding within a realistic cosmological lightcone is
important for intensity mapping, since a spectral channel at observed
wavelength $\lambda_{\rm obs}$ receives continuum emission from galaxies
over a broad range of redshifts, while different rest-frame emission
lines enter the same channel at different redshifts. We therefore use all
available redshift partitions when constructing the main SPHEREx-like
maps, applying redshift cuts only for diagnostic tests. With a dark matter particle mass of
$M_{\rm DM}\sim9.6\times10^{7}\,M_\odot\,h^{-1}$, the mass threshold
for resolved haloes is set to $M_{\rm res}\sim10^{10}\,M_\odot$, corresponding to
roughly 100 dark matter particles. Galaxy formation becomes quite inefficient at masses smaller than this halo mass \citep{Somerville2002}, and therefore we do not expect lower mass halos to contribute significantly to the LIM signal. 

The lightcone table provides the sky position and redshift of each galaxy;
the SAM property table provides quantities used in the line and dust
models, including star-formation rate, metallicity, and dust parameters;
and the photometry table provides observed-frame continuum fluxes.
The tables are matched using the galaxy identifiers supplied with the
catalog, and the matches are checked by comparing redshifts across the
linked files. The SAM catalogs do not include nebular emission, so we add the line emission for each galaxy, based on its properties, as described in Section~\ref{sec:linemodel}.
Nebular attenuation is then applied using dust properties from the
catalog, treating the attenuation of the emission lines separately from
the stellar-continuum attenuation already included in the photometry.
This allows dust attenuation to vary across the galaxy population and across lines,
rather than entering as a single global correction factor in the modeling framework. Furthermore, the continuum maps are built from the dust-attenuated SAM photometry. In
our analysis this continuum component acts as the bright, spectrally
smooth foreground against which the line intensity fluctuations must be
recovered. Because the continuum and line emission are assigned to the
same underlying galaxy population, the mock maps preserve the spatial relationship
between continuum emission, line emission, dust attenuation, and galaxy
clustering \citep{Yung2023}. Finally, the catalog is projected onto a regular angular and spectral
grid, pixelized to $1'$ resolution. 

In Figure~\ref{fig:sam_lightcone}, we visualize the geometry of the SAM
lightcone catalogue. The full lightcone contains $1.36\times10^7$ galaxies
spanning a redshift range of approximately
$0<z<10$. The horizontal axis shows redshift, while the vertical axis shows
one transverse comoving coordinate relative to the central line of sight.
The widening of the distribution with redshift is a geometric effect: for a
fixed angular footprint, the corresponding transverse comoving size grows
with distance. In the plotted sample, the lightcone reaches transverse
offsets of order $\pm 110~{\rm cMpc}$ at the high-redshift end.
Points are colored by host halo mass, illustrating that the most massive
haloes are preferentially found at lower redshift, while such systems become
increasingly rare at early cosmic times. For visual clarity, we subsample the catalogue and show approximately $1\%$ of the full lightcone population in the plot.

Figure~\ref{fig:samprops} summarizes the redshift support of the SAM lightcone and the modeled rest-frame optical lines. The histogram shows the galaxy redshift distribution of the catalog. The galaxy counts peak near $z\simeq2.2$ and decline toward both lower and higher redshift, falling steadily beyond the peak out to the catalog limit at $z\simeq10$. This peak occurs near the epoch of maximum cosmic star-formation activity, although the detailed shape of the histogram also reflects the lightcone volume and catalog construction. The horizontal bars mark the redshift intervals over which each of the four modeled lines falls within the $0.75$--$5\,\mu{\rm m}$ observed-frame wavelength range: \ha\ over $0.14\lesssim z\lesssim6.62$, \hb\ over $0.54\lesssim z\lesssim9.29$, \oiii\ over $0.50\lesssim z\lesssim8.99$, and \oii\ over $1.01\lesssim z\lesssim10$. The shaded region marks the interval $1.0\lesssim z\lesssim6.6$, where all four modeled lines are simultaneously observable. This overlap sets the redshift range in which all six same-redshift line-pair ridges can contribute to the channel-correlation matrices.

\subsection{Line Luminosity Model}
\label{sec:linemodel}

Although hydrogen produces a rich recombination spectrum, restricting our
attention to \ha\ and \hb\ costs us very little. The Balmer line strengths fall
steeply with upper level under Case B conditions ($T_e \simeq 10^4$~K,
$n_e \sim 10^{2}$--$10^{4}~{\rm cm}^{-3}$), with
${\rm H}\gamma/\hb \simeq 0.47$ and ${\rm H}\delta/\hb \simeq 0.26$
\citep{Osterbrock-2006-book}, and since every Balmer line traces the same ionizing
photon budget, the higher members of the series add flux at the few percent
level without adding any new physical information. The Paschen and Brackett
series are weaker still, ${\rm Pa}\alpha/\ha \simeq 0.12$ and
${\rm Br}\gamma/\ha \simeq 0.01$ in Case B, and their long rest wavelengths
($1.875$ and $2.166~\mu{\rm m}$) push them out of a $0.75$--$5~\mu{\rm m}$
bandpass beyond $z \simeq 1.7$, where the comoving volume available for
intensity mapping is modest. ${\rm Pa}\alpha$ does suffer less dust
attenuation than \ha, but this does not make up for an order of magnitude
deficit in intrinsic luminosity. \lya\ is a different story: it is
intrinsically the brightest hydrogen line, with
${\rm Ly}\alpha/\ha \simeq 8.7$ in Case B, but it only enters the SPHEREx
bands at $z \gtrsim 5.2$, beyond the range we consider, and resonant
scattering makes its escape fraction both low and highly stochastic, so it
does not trace star formation in the clean way the Balmer lines do. For our
purposes, \ha\ and \hb\ together capture essentially all of the recoverable
hydrogen line signal.

We are interested in the four rest-frame optical lines, the
strongest sources of emission accessible to a SPHEREx-like
$0.75$--$5~\mu{\rm m}$ survey over the relevant redshift range: the Balmer
recombination lines \ha\ and \hb, and the singly and doubly ionized oxygen
forbidden lines \oii\ and \oiii. The rest wavelengths we adopt are
\begin{equation}
  \lambda_{\rm rf} = (0.6563,\,0.4861,\,0.5007,\,0.3727)~\mu{\rm m}
\end{equation}
for \ha, \hb, \oiii, and \oii, respectively. These four lines are not
interchangeable tracers, since they form through different processes and
therefore respond differently to the star-formation rate, the gas-phase
metallicity, and the ionization state of the interstellar medium; it is this
diversity that lends diagnostic power to the multi-line cross-channel
spectrum.

The Balmer lines are produced by recombination in H\,\textsc{ii} regions.
Ultraviolet photons from hot, young, massive stars ionize the surrounding
hydrogen, and the freed electrons subsequently recombine and emit
H$\alpha$ and H$\beta$ through the ensuing radiative cascade
\citep{Osterbrock-2006-book}. Because the recombination rate tracks the
rate at which ionizing photons are produced, and because short-lived massive
stars dominate that ionizing budget, the Balmer luminosities are close to
instantaneous tracers of the star-formation rate, with only a weak
dependence on metallicity \citep{Kennicutt1998}. In contrast, the oxygen lines arise from collisionally excited forbidden
transitions. Their luminosities are therefore sensitive not just to the
ionizing photon rate and the oxygen abundance, but also to the temperature
and ionization parameter of the emitting gas, and so they carry information
about chemical enrichment and the hardness of the radiation field rather
than the star-formation rate alone. This shapes the model that follows: we
anchor all four lines to the SFR at lowest order, while allowing that the
oxygen normalizations also depend implicitly on gas properties.

We take the line luminosities to be linear in SFR,
\begin{equation}
  L_i = r_i\left(\frac{\rm SFR}{M_\odot~{\rm yr}^{-1}}\right),
  \label{eq:line_sfr}
\end{equation}
where $r_i$ has units of ${\rm erg~s^{-1}}\,(M_\odot~{\rm yr}^{-1})^{-1}$.
This is a well-motivated choice for star-formation tracers, and especially
for the Balmer lines, as noted above. It is an effective,
population-averaged relation that folds the metallicity and
ionization-parameter dependence into a single coefficient. Motivated by the
calibration of \citet{Kennicutt1998}, the emission-line measurements of
\citet{Ly2007}, and the SPHEREx line intensity-mapping modeling of
\citet{Gong2014,Gong2017}, we adopt
\begin{align}
  r_{\ha}   &= 1.27\times10^{41},\\
  r_{\hb}   &= 0.44\times10^{41},\\
  r_{\oiii} &= 1.32\times10^{41},\\
  r_{\oii}  &= 0.71\times10^{41}.
\end{align}
The \ha\ coefficient is the canonical Kennicutt value relating
\ha\ luminosity to the star-formation rate for a fully sampled initial
mass function. The remaining coefficients are fixed relative to
\ha\ through the line ratios characteristic of star-forming galaxies, so
that the model reproduces the relative strengths of the four lines
rather than treating their normalizations as independent free
parameters.

\begin{figure*}[htbp]
    \centering
    \includegraphics[width=\linewidth]{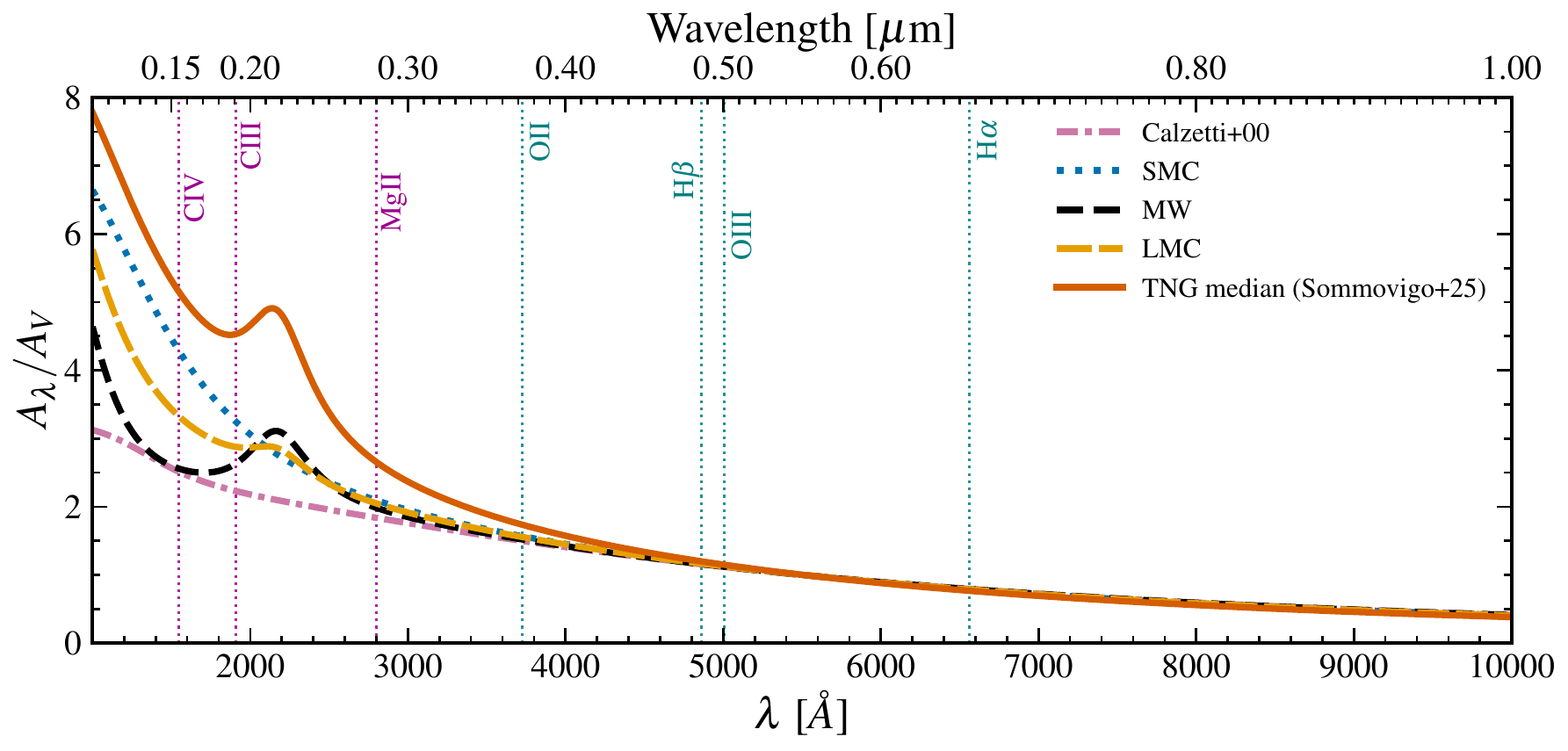}
\caption{
Attenuation curves used to illustrate the wavelength dependence of dust
suppression. The curves show $A_\lambda/A_V$ as a function of rest-frame
wavelength for several commonly used prescriptions: Calzetti, SMC, Milky
Way, LMC, and the TNG median curve from \citet{Sommovigo2025}. Teal dotted lines mark the rest-frame wavelengths of
the four emission lines in our LIM model, while magenta dotted lines
mark prominent UV lines for reference. The prescriptions agree closely
redward of $\sim4000$\,\AA\ but diverge strongly in the ultraviolet,
where features such as the 2175\,\AA\ bump appear. Because
$A_\lambda/A_V$ rises toward shorter wavelengths, blue lines such as
\oii\ and \hb\ are attenuated more strongly than \ha.
}
    \label{fig:attenuation_curves}
\end{figure*}

The \hb\ normalization corresponds to a fixed ratio
$L_{\hb}/L_{\ha}\simeq0.35$, set by the atomic physics of hydrogen
recombination. Under Case B conditions, appropriate for the optically
thick H\,\textsc{ii} regions of star-forming galaxies, the intrinsic
\ha/\hb\ ratio is close to $2.86$ for typical electron temperatures and
densities, and the value adopted here is consistent with that
expectation and with the ratio used in the line intensity-mapping
literature \citep{Osterbrock-2006-book, Gong2017}. Fixing this ratio is
important for the present study, because the \ha/\hb\ Balmer decrement is frequently used as an observational probe of attenuation, and our dust treatment
in Sec.~\ref{sec:dustmodel} is applied on top of these intrinsic normalizations. The oxygen coefficients
likewise place \oiii\ slightly above \ha\ and \oii\ at roughly half the
\ha\ value, in line with the typical strengths of these lines in
star-forming galaxies at typical metallicities in the redshift interval $z\sim0$--$5$.

There are two simplifying assumptions in this baseline model that should be
noted explicitly, since they affect how the cross-channel spectrum is
interpreted. First, we hold the coefficients $r_i$ fixed for the entire
galaxy population, rather than allowing the oxygen line ratios to vary with
ISM properties that may correlate with stellar mass, metallicity, ionization
parameter, and redshift \citep[see e.g.][]{Kewley2004, Tremonti2004,
Kewley2013, Sanders2021}. The \oii\ luminosity in particular is known to be
highly sensitive to metallicity \citep[e.g.][]{Kewley2004}, and even the
conversion from SFR to Balmer luminosity carries a mild dependence on the
underlying stellar-population modeling, including metallicity
\citep[e.g.][]{KennicuttEvans2012}. Because we hold $r_i$ fixed, this
galaxy-to-galaxy variation is absent from our model; including it would
reduce the contrast of the same-redshift ridges and add stochasticity to the
cross-channel spectra. Second, we account only for line emission powered by
star formation and neglect AGN line emission. AGN narrow-line regions can
contribute appreciably to the \oiii\ emission, which would break the simple
linear relation with SFR.

These choices are deliberate: they provide a clean demonstration of how the
other components of the survey affect the signal and its recovery, isolating
the effects of dust, continuum, and survey geometry. A transparent,
population-averaged relation between line strengths is what this kind of
experiment requires. The SAM catalog of Sec.~\ref{sec:sam_catalog} provides
the star-formation rate, metallicity, and dust content of each galaxy
individually, so metallicity-dependent line ratios can be introduced in
future work without any further structural changes to our model.

\subsection{Dust Attenuation}
\label{sec:dustmodel}
 
In the Santa Cruz model, the dust attenuation of each galaxy is estimated based on its cold gas mass, gas-phase metallicity,
and disc size \citep{Somerville2012,Yung2019}; thus the optical depth is based on properties that are physically linked to those that govern the line emission. This self-consistency matters for the
present work as dusty galaxies are preferentially metal enriched and
star-forming, so attenuation does not act as a random dimming but as a
weight that is correlated with the line luminosities themselves.

The SAM lightcone catalog provides the face-on $V$-band optical depth for
each galaxy, computed as described above. To compute the attenuation of the
nebular line emission, we use this catalog optical-depth parameter as an
effective face-on $V$-band optical depth and include inclination dependence
through the same slab geometry adopted in the SAM \citep{DeLucia2006},
\begin{equation}
    \tau_{V,g}
    =
    { \tau_{0,g} \over \max(|\cos i_g|,\cos i_{\rm min}) },
    \qquad
    \cos i_{\rm min}=0.05 .
    \label{eq:tauv_sam}
\end{equation}
The $1/\cos i$ factor accounts for the longer path length through the dust
disc as a galaxy is viewed away from face-on, and the floor
$\cos i_{\rm min}=0.05$ prevents the optical depth from diverging for
systems that are almost exactly edge-on, where the thin-disc approximation
underlying the geometry breaks down in any case.

For self-consistency with the stellar-continuum photometry, which the SAM
computes under a mixed dust-star slab geometry, we adopt the same slab model
to convert this optical depth into the reference stellar-continuum
attenuation \citep{Somerville2012,Yung2019},
\begin{equation}
    A_{V,g}^{\rm star}
    =
    -2.5\,\log_{10}\!\left[
    \frac{1-\exp(-\tau_{V,g})}{\tau_{V,g}}
    \right].
    \label{eq:slab}
\end{equation}
Unlike a foreground screen, the slab attenuation saturates at high optical
depth, since the observed emission becomes dominated by the least obscured
material along the line of sight. This bounded behavior makes an explicit
attenuation cap unnecessary and ensures that the line and continuum
components are attenuated under the same physical geometry.

The nebular lines are generally expected to be attenuated more
strongly than the stellar continuum, if the line emitting gas is more
closely associated with dusty star-forming regions than the older
stellar population \citep{Calzetti2000, CharlotFall2000}. We model
this using
\begin{equation}
    A^{\rm neb}_{V,g} = f_{\rm neb}\, A^{\rm star}_{V,g},
    \qquad f_{\rm neb} = \frac{1}{0.44} = 2.27,
    \label{eq:fneb}
\end{equation}
following the relation between nebular and stellar color excesses
\citep{Calzetti2000, Reddy2020}. This prescription preserves the
galaxy-to-galaxy variation in dust attenuation, based on the galaxy
physical properties, while including the observed average offset
between line and continuum attenuation.

The attenuation at the wavelength of line $X$ is then obtained by
evaluating the nebular attenuation curve at the rest-frame wavelength
of the line,
\begin{equation}
    A^{\rm neb}_{X,g}
    = R_{\rm att}(\lambda_{X,\rm rest})\, A^{\rm neb}_{V,g},
    \qquad
    R_{\rm att}(\lambda) \equiv \frac{A_\lambda}{A_V},
    \label{eq:Aline}
\end{equation}
so that the line attenuation depends on both the \ion{H}{2} region
dust column, through $A^{\rm neb}_{V,g}$, and the rest-frame
wavelength of the line, via the (nebular) attenuation curve
$R_{\rm att}(\lambda)$, normalized such that
$R_{\rm att}(\lambda_V) = 1$ at $\lambda_V = 0.55\,\mu$m. The
attenuation curve depends primarily on the dust composition and
star--dust geometry \citep[e.g.][]{SalimNarayanan2020,
Sommovigo2025}, and attenuation tends to be stronger at bluer
wavelengths. Therefore, shorter wavelength lines such as [\ion{O}{2}]
and H$\beta$ are expected to have smaller transmission factors than
H$\alpha$ for the same $A^{\rm neb}_{V}$.

Our fiducial choice for $R_{\rm att}$ is the Calzetti+00 attenuation curve
from \citet{Sommovigo2025}, written in the \citet{Li2008} parameterization as
\begin{align}
R_{\rm att}(\lambda)
=&\,
{c_1 \over
(\lambda/0.08)^{c_2} + (0.08/\lambda)^{c_2} + c_3}
\nonumber\\
&+
{233\left[
1 -
{c_1 \over 0.145^{-c_2}+0.145^{c_2}+c_3}
-
{c_4 \over 4.60}
\right]
\over
(\lambda/0.046)^2 + (0.046/\lambda)^2 + 90}
\nonumber\\
&+
{c_4 \over
(\lambda/0.2175)^2 + (0.2175/\lambda)^2 - 1.95},
\label{eq:attenuation_curve}
\end{align}
where $\lambda$ is in microns. For the Calzetti+00 curve we use $(c_1,c_2,c_3,c_4)=(44.9,\,7.56,\,61.2,\,0)$.

With the line attenuation in hand, we define the dust transmission factor for line $X$ in galaxy $g$ as
\begin{equation}
    f_{{\rm dust},X,g}
    \equiv
    {L^{\rm obs}_{X,g} \over L^{\rm int}_{X,g}}
    =
    10^{-0.4 A^{\rm neb}_{X,g}} .
    \label{eq:fdust_def}
\end{equation}
Thus $f_{{\rm dust},X,g}=1$ corresponds to no attenuation, while
$f_{{\rm dust},X,g}<1$ indicates suppression of the intrinsic line
luminosity by dust.

\begin{figure}[htbp]
    \centering
    \includegraphics[width=\linewidth]{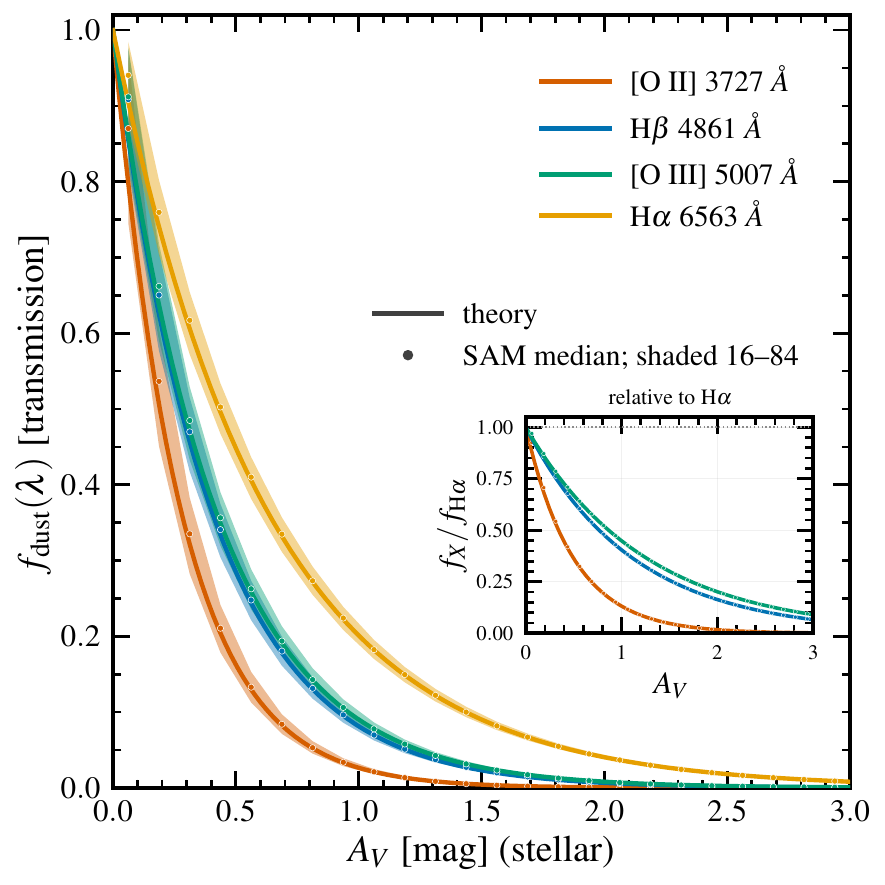}
 \caption{
    Dust transmission for the four rest-frame optical lines used in this
    work, shown as a function of the stellar-continuum attenuation
    $A_V^{\rm star}$. Solid curves show the model prediction for the
    Calzetti attenuation curve with the nebular enhancement
    $A_V^{\rm neb}=2.27\,A_V^{\rm star}$. Points show the median
    transmission for star-forming galaxies in the selected SAM redshift
    slice, binned by $A_V^{\rm star}$, and shaded regions show the
    1$\sigma$ uncertainty. For a given $A_V^{\rm star}$, \oii\ is suppressed
    most strongly and \ha\ least, while \hb\ and \oiii\ lie so close in
    wavelength that their transmissions nearly coincide. The inset shows
    the transmission of \oii, \hb, and \oiii\ relative to \ha,
    illustrating the differential attenuation that changes the observed
    line ratios across the galaxy population.
    }
    \label{fig:fdust_vs_av}
\end{figure}

In Figure~\ref{fig:attenuation_curves} we show the wavelength dependence of
$R_{\rm att}(\lambda)=A_\lambda/A_V$ for the attenuation curves
considered here, with vertical markers indicating the rest-frame
wavelengths of the emission lines used in the LIM maps. The figure
illustrates why dust cannot be represented as a single multiplicative
factor: \oii\ and \hb\ lie at shorter wavelengths and are therefore more
strongly attenuated, while \ha\ lies at a longer wavelength and is less
suppressed. The differences among the curves are largest in the
ultraviolet and blue optical, where features such as the
$2175~\text{\AA}$ bump and the slope of the curve are most pronounced.
Both effects act on the line ratios: dust suppresses \hb\ and \oii\
relative to \ha\ for any curve, and the choice of curve sets the size of
that suppression. The resulting shifts in ratios such as $\ha/\hb$ and
$\ha/\oii$ propagate directly into the relative amplitudes of the line
intensity maps built from each transition.

At the map level, this enters as a galaxy-dependent weight. Writing
the transmission as $T_{X,g} \equiv f_{{\rm dust},X,g}$, a schematic line map is
\begin{equation}
I_X(\hat{\mathbf{n}},\nu_i) = \sum_g
\frac{L^{\rm int}_{X,g}\, T_{X,g}}
     {4\pi D_L^2(z_g)\, \Delta\nu_i\, \Delta\Omega_{\rm pix}}
\, W_i\!\left[\nu_X/(1+z_g)\right]
\Pi_{\rm pix}(\hat{\mathbf{n}}-\hat{\mathbf{n}}_g),
\label{eq:linemap}
\end{equation}
where $W_i$ is the spectral bandpass of channel $i$ evaluated at the
redshifted line frequency, and $\Pi_{\rm pix}$ is the pixel window
function assigning galaxy $g$ at position $\hat{\mathbf{n}}_g$ to sky
pixels. Because the transmission enters as a multiplicative weight on
each source, the large-scale clustering and shot noise terms are both
modified. For
example, the shot contribution to a cross-spectrum between lines $X$ and
$Y$ contains the weighted product
$L^{\rm int}_{X,g}L^{\rm int}_{Y,g}T_{X,g}T_{Y,g}$.  Since $T_{X,g}$ is
correlated with the galaxy properties that also determine line luminosity,
a global rescaling cannot reproduce the full effect of dust on the LIM
statistics. For the stellar continuum maps we use the dust-attenuated SAM photometry
provided by the catalog.  Thus the continuum and line components are
attenuated consistently in the sense that both are tied to the same galaxy
population and dust optical depth information, while still allowing the
nebular emission to receive the additional attenuation appropriate for
H\,\textsc{ii} regions.  We also construct no-dust line maps by setting
$T_{X,g}=1$.  Comparing the dusty and no-dust maps isolates the impact of
attenuation on the line intensities, cross-channel spectra, and normalized
correlation matrices before continuum cleaning is applied.

Figure~\ref{fig:fdust_vs_av} illustrates the resulting transmission for
the four optical lines used in the LIM maps. For the same
$A_V$, \oii\ is suppressed most strongly, while \ha\ remains
the least attenuated because of its longer rest-frame wavelength. At $A_V=1$, for example, \ha\ retains roughly
$\sim18\%$ of its flux while \oii\ retains only $\sim5\%$. By
contrast, \hb\ and \oiii\ lie so close in wavelength that their
transmission curves nearly coincide, so dust leaves the \hb/\oiii\
ratio almost unchanged even as it strongly suppresses both lines. The
SAM points trace the theoretical curves by construction, since each
galaxy's transmission is computed from its SAM-derived
$A_V$; the shaded scatter reflects the finite bin width. The
inset highlights the quantity most relevant for multi-line analyses:
dust changes the transmission of \oii, \hb, and \oiii\ relative to
\ha\ by amounts that differ from line to line, so the effect cannot be
represented by a single line intensity-dependent amplitude correction.
Conversely, line pairs with similar rest wavelengths, such as
\hb\ and \oiii, provide nearly dust-independent ratios that can serve
as anchors when interpreting the dust-sensitive combinations.

\section{Experiments and Observables}
\label{sec:experiments}
In this section, we describe how the modeled emission lines from galaxies are mapped into an observable
spectral data cube. We first specify the observational geometry and the
SPHEREx-like wavelength channel configuration used in this work. We then
define the cross-channel angular power spectra and the corresponding
normalized correlation matrices, which form the primary summary statistics
used in the remainder of the analysis.

\subsection{General Setup}
\label{sec:general_setup}

The forward model developed in this work applies to low-resolution
spectral imaging experiments that map the sky in many contiguous
observed-frame wavelength channels. Such an experiment may be
characterized by its wavelength coverage, spectral resolving power
$R=\lambda/\Delta\lambda$, angular resolution, and survey footprint. The
basic observable is a spectral data cube,
\begin{equation}
    I_{\nu,i}({\bm \theta}),
\end{equation}
the specific intensity in channel $i$ as a function of sky position
${\bm \theta}$ at an observed frequency $\nu$. Each observed channel receives contributions from stellar
continuum emission over a broad range of redshifts and from multiple rest-frame emission
lines redshifted into that channel. Thus an individual channel is not a
single redshift slice, but a superposition of continuum and line emission
from different redshifts.
The goal of line intensity mapping is to recover the statistical
properties of the line emitting large-scale structure from this mixed
spectral cube. 

We focus on the angular auto- and cross-power spectra
between wavelength channels. We refer to the two-dimensional array $C_\ell(\lambda_i,\lambda_j)$, indexed
by all pairs of observed wavelength channels, as the channel-pair
power-spectrum matrix. At each multipole $\ell$ it is a symmetric
$N_\lambda \times N_\lambda$ matrix whose diagonal elements are the
auto-spectra of individual channels and whose off-diagonal elements measure
the cross-correlation between the intensity fluctuations in two different
channels. A pair of channels $(\lambda_i,\lambda_j)$ acquires a large
off-diagonal entry when the two channels receive emission from the same
comoving volume, most notably when two different rest-frame lines emitted
at a common redshift satisfy $\lambda_i/\lambda_{a,\rm rest} =
\lambda_j/\lambda_{b,\rm rest} = 1+z$, which is what produces the
same-redshift ridges discussed below.

Although the formalism is experiment-independent, we adopt a
SPHEREx-like spectral imaging survey as our default
example \citep{Dore2018}. SPHEREx provides a natural benchmark because its wavelength
coverage and low spectral resolution are well matched to optical and
near-infrared LIM studies. In practice, our maps are built from the
SAM lightcone catalog and should be interpreted as controlled
SPHEREx-like mock observations of that footprint, rather than as a full
SPHEREx survey simulation. We do not model the map-making step, in which individual exposures are mosaicked into wide-field maps; for SPHEREx this introduces correlated noise, position-dependent spectral response, and large-scale gain variations that would enter the analysis through the same mode
coupling and transfer function treatment applied here. Our maps begin
from an idealized pixelized cube, so the transfer functions we derive
capture cleaning-induced losses only.

We work in surface-brightness units of ${\rm Jy\,sr^{-1}}$ throughout and
the angular power spectra defined in Section~\ref{sec:cl}
accordingly carry units of $({\rm Jy\,sr^{-1}})^2\,{\rm sr}$, since the
power spectrum is the variance of the intensity field per unit area in
multipole space and therefore scales as the squared field amplitude
times solid angle. A line with rest wavelength
$\lambda_{\rm rest}$ observed in channel $i$ originates from redshift
$z=\lambda_i/\lambda_{\rm rest}-1$, so each channel mixes several
discrete redshift slices of line emission with the continuous redshift
integral of the continuum. Instrumental noise and other survey
systematics are not included in the baseline maps; we introduce a noise
model where needed for the inference tests in
Section~\ref{sec:inference}.
\subsection{SPHEREx-like Spectral Channels}
\label{sec:channels}

We construct a simplified SPHEREx-like channel grid with 91 observed-frame
channels spanning $0.75$--$5.0~\mu{\rm m}$. The channel centers are fixed
in observed wavelength, and each channel is assigned an effective width
$\Delta\nu_i$ from the channel edges. This approximation captures the
essential property needed for this work: a contiguous low-resolution
near-infrared spectral cube. Instrumental details such as detector noise,
zodiacal emission, detailed bandpasses, and survey masks are not included
in the baseline maps and optional beam smoothing is applied separately when
specified.

The total intensity in each channel is written as the sum of continuum
and line components,
\begin{equation}
  I_{\nu,i}({\bm \theta}) =
  I_{\nu,i}^{\rm cont}({\bm \theta}) +
  I_{\nu,i}^{\rm line}({\bm \theta}) ,
  \label{eq:totalmap}
\end{equation}
where $I_{\nu,i}^{\rm cont}$ is the stellar-continuum contribution from
galaxies at all redshifts, and $I_{\nu,i}^{\rm line}$ is the contribution
from the four modeled lines redshifted into channel $i$.

The continuum cube is obtained by interpolating each galaxy's
observed-frame SAM photometry to the SPHEREx-like channel centers and
binning the resulting flux density into angular pixels. For a pixel of
solid angle $\Omega_{\rm pix}$,
\begin{equation}
  I_{\nu,i}^{\rm cont}({\bm \theta}) =
  \frac{1}{\Omega_{\rm pix}}
  \sum_{g\in{\rm pix}} f_{\nu,g}(\lambda_i),
  \label{eq:contmap}
\end{equation}
where $f_{\nu,g}(\lambda_i)$ is the interpolated observed-frame flux
density of galaxy $g$ in channel $i$. This component represents the
bright, spectrally smooth continuum against which the line fluctuations
must be recovered.

For an emission line with luminosity $L_{\rm line}$, the integrated
observed flux is
\begin{equation}
  F_{\rm line} = \frac{L_{\rm line}}{4\pi D_L^2},
  \label{eq:lineflux}
\end{equation}
where $D_L$ is the luminosity distance. The observed line wavelength is
$\lambda_{\rm obs}=\lambda_{\rm rest}(1+z)$. When using a Gaussian
spectral profile, we assign the observed-frame line width
\begin{equation}
  \sigma_\lambda =
  \frac{\lambda_{\rm obs}}{2.355\,R},
  \label{eq:gaussline}
\end{equation}
and integrate the normalized Gaussian over the channel edges. If
$p_{ig}$ is the fraction of line flux from galaxy $g$ falling in channel
$i$, then the corresponding channel-averaged flux density is
\begin{equation}
  f_{\nu,i,g}^{\rm line} =
  \frac{F_{{\rm line},g}p_{ig}}{\Delta\nu_i}.
  \label{eq:linejy}
\end{equation}
The line intensity map is then obtained by summing
$f_{\nu,i,g}^{\rm line}/\Omega_{\rm pix}$ over all galaxies in the pixel.
Because the continuum and line emission are assigned to the same galaxies,
the mock cube preserves the spatial correlation between the line signal
and the continuum.

\begin{figure*}[t]
    \centering    \includegraphics[width=0.97\textwidth]{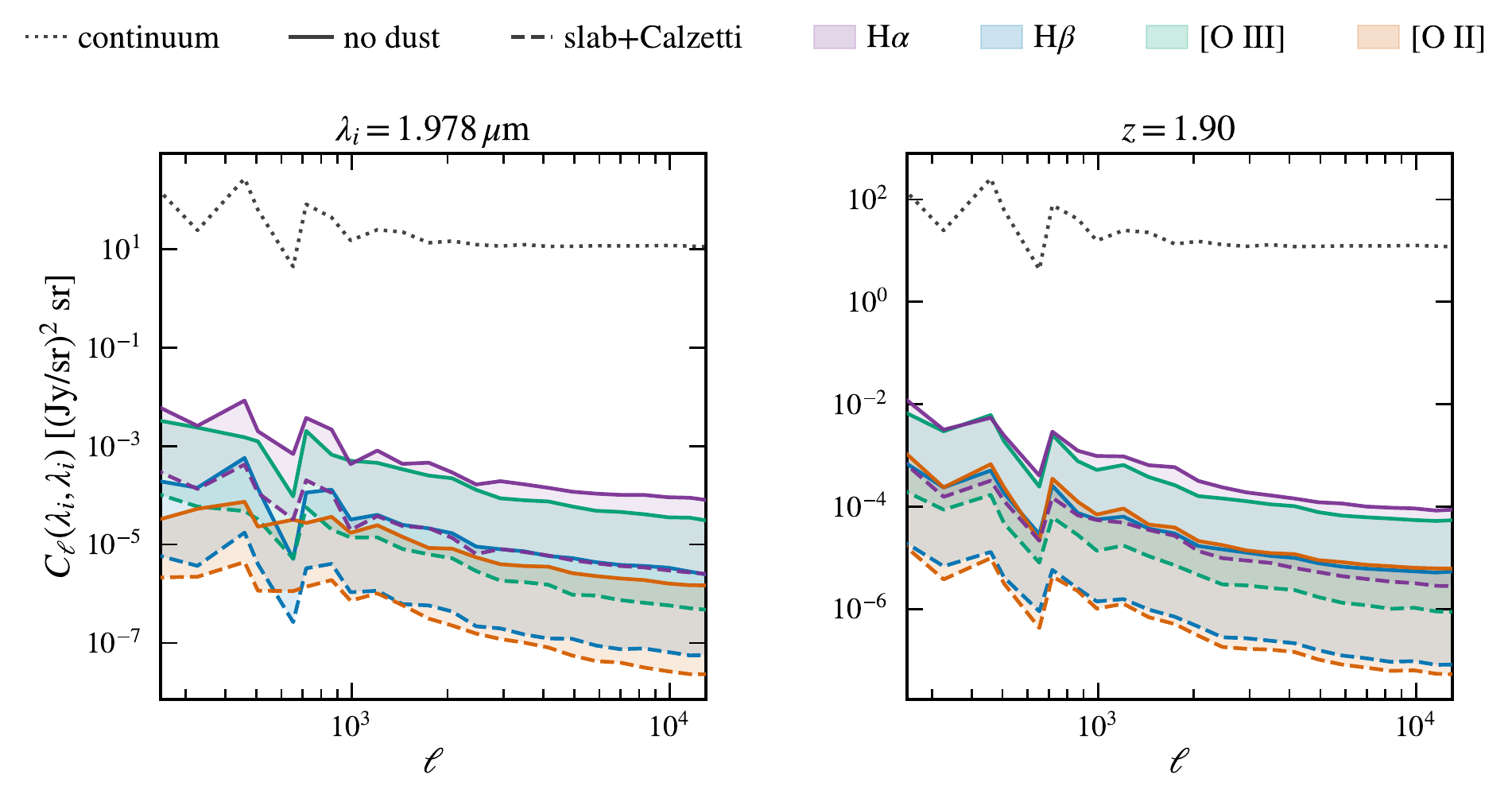}
    \caption{
    Angular power spectra as a function of nebular dust attenuation in the line
    maps. The dotted gray curve shows the continuum-only auto-spectrum, which is
    held fixed in the slab comparison. Solid colored curves denote the
    no-dust line spectra, while dashed colored curves denote the corresponding
    slab attenuated spectra. Shaded bands connect each no-dust line
    spectrum to its attenuated counterpart. The left panel compares the components
    in a single observed channel, $\lambda_i = 1.978\,\mu{\rm m}$. The right panel
    fixes the emission redshift to $z = 1.90$, so that each line is evaluated in
    the corresponding observed channel.
}
    
    \label{fig:fdust-cl-bands}
\end{figure*}

\subsection{Angular Power Spectra and Correlation Matrices}
\label{sec:cl}

For each channel map, we subtract the mean intensity and estimate
flat sky Fourier modes on the pixelized map. The binned cross-power
spectrum between channels $i$ and $j$, with wavelengths $\lambda_i$ and $\lambda_j$, is
\begin{equation}
  C_b^{ij} =
  \frac{A}{N_b}
  \sum_{{\bm \ell}\in b}
  \tilde I_i({\bm \ell})
  \tilde I_j^*({\bm \ell}),
  \label{eq:cl}
\end{equation}
where $A$ is the map area, $N_b$ is the number of Fourier modes in the
multipole bin, and $\tilde I_i({\bm \ell})$ is the Fourier transform of
the mean-subtracted map. The diagonal elements $C_\ell^{ii}$ are channel
auto-spectra, while the off-diagonal elements $C_\ell^{ij}$ contain the
cross-channel information.

We additionally use the normalized correlation matrix of channels as,
\begin{equation}
  r_\ell^{ij} =
  \frac{C_\ell^{ij}}{\sqrt{C_\ell^{ii}C_\ell^{jj}}}.
  \label{eq:rij}
\end{equation}
This isolates the spectral coherence between
channels from their amplitudes. For example, $|r^{ij}_\ell|\le1$, with values near
unity indicates that the two channels trace the same underlying
structure, and values near zero refers to statistically independent
fluctuations. Because the normalization removes channel-level amplitude
effects, $r^{ij}_\ell$ is not sensitive to overall calibration and to any
attenuation that acts uniformly on a channel pair, making it a
complementary statistic to the dimensional $C^{ij}_\ell$. The morphology of the $r^{ij}_\ell$ matrix is itself diagnostic as the spectrally smooth
continuum produces broad regions of $r\simeq1$ but correlated line
pairs produce localized same-redshift ridges.

Our baseline estimator is the pseudo-$C_\ell$ method implemented in
\textsc{NaMaster} \citep{Hivon2002, Alonso2019}. Although our mock
fields are small and nearly rectangular, they are still finite and the map
edge acts similar to a mask, coupling angular modes that mixes power across
multipoles. The pseudo-$C_\ell$ approach deals with this directly,
computing the mode coupling matrix from the footprint geometry and
correcting the binned bandpowers, rather than assuming the periodic
boundaries that a simple FFT requires. 
A SPHEREx-like measurement comes with masks and weights
variations that enter through the same mode coupling treatment,
so building the pipeline on \textsc{NaMaster} allows us to move from mocks to data. We still compare
against a simple flat-sky FFT estimator, which is fast and easy to
interpret, and for the weakly masked fields used here the two agree
well. We therefore use \textsc{NaMaster} throughout and keep the FFT as
a consistency check.

We show how nebular dust attenuation changes the line power spectrum in
Figure~\ref{fig:fdust-cl-bands}.  In this comparison, the continuum-only spectrum is
held fixed, while the line spectra are shown both without nebular attenuation
and with the slab-attenuation model.  Solid colored curves show the
no-dust line spectra, and dashed colored curves show the corresponding
slab attenuated spectra.  In the fixed observed-wavelength panel, all
components are evaluated in the same SPHEREx-like channel, so the comparison
shows the range of angular power that would be measured in a single wavelength
slice.  The continuum remains several orders of magnitude brighter than the
individual line powers over most multipoles, reflecting the broadband nature
of the galaxy continuum.  The line spectra are reduced by the slab
attenuation, with the strongest suppression for the shorter-wavelength lines.
The fixed-redshift panel gives a complementary view.  Instead of forcing all
lines into the same observed channel, each transition is evaluated at the
wavelength corresponding to $z\simeq1.9$, near the peak of cosmic star
formation, so the comparison is tied to the physical line-emitting population
at a single epoch.

\begin{figure*}[htbp]
    \centering
    \includegraphics[width=\linewidth]{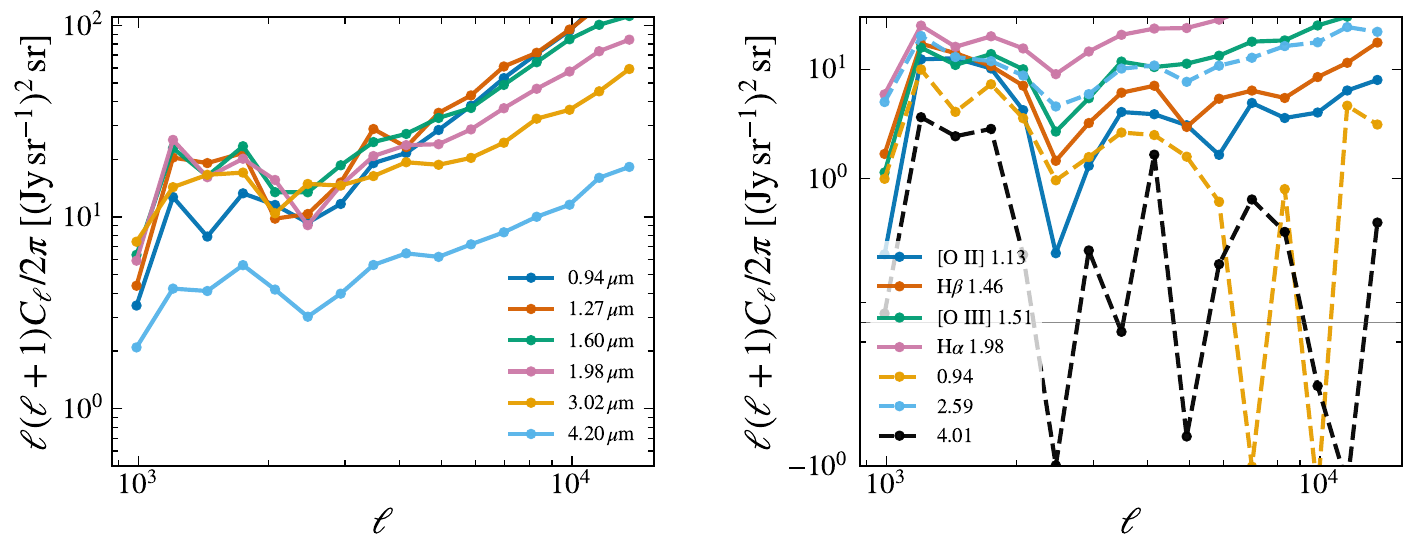}
    \caption{ Left: auto-spectra
    $D_\ell^{ii}=\ell(\ell+1)C_\ell^{ii}/2\pi$ for representative observed
    wavelength channels.  Right: cross-spectra between an \ha-selected
    anchor channel at $\lambda_i=1.98\,\mu{\rm m}$, corresponding to
    $z\simeq2.02$, and several other observed wavelength channels
    $\lambda_j$.  Solid curves mark channels corresponding to the
    same-redshift \oii, \hb, \oiii, and \ha\ lines for this anchor
    redshift; dashed curves show comparison wavelengths that are not chosen
    to lie on the same-redshift line-pair loci.  The cross-spectrum panel
    uses a symmetric logarithmic vertical scale so that positive and
    negative estimates can be shown together.  The figure illustrates both
    the scale dependence of the line power and the enhanced cross-power
    expected when two channels trace the same comoving volume through
    different emission lines.
    }
    \label{fig:cl_ell_spectra}
\end{figure*}

\section{Results}
\label{sec:results}

With the mock maps and estimator pipeline in place, in this section we present the
multi-line signal before any continuum cleaning is applied. We begin
with the angular power spectra of the line-only maps
(Section~\ref{sec:power_spec_analysis}), then turn to the cross-channel
correlation matrices and same-redshift ridges that form our central
observable (Section~\ref{sec:line_only_correlations}), the impact of
dust on observables (Section~\ref{sec:dust_vs_nodust}), and the decomposition of cross-channel power
into clustering and shot noise (Section~\ref{sec:shot_noise}).

\subsection{Power Spectrum Analysis}
\label{sec:power_spec_analysis}

The main observable in our analysis is the cross-channel angular
power spectrum, which can be expressed as
\begin{equation}
    C_\ell^{ij}
    \equiv
    C_\ell(\lambda_i,\lambda_j),
    \label{eq:clij_results}
\end{equation}
where $i$ and $j$ label the observed wavelength channels.  The diagonal
elements, $C_\ell^{ii}$, are auto-spectra of individual observed channels,
while the off-diagonal elements measure cross-correlations between
different wavelengths.  This distinction is central for MLIM:
two different observed channels can trace the same large-scale structure
when they correspond to different rest-frame emission lines emitted at
the same redshift.

\begin{figure*}[htbp]
    \centering
    \includegraphics[width=0.9\textwidth]{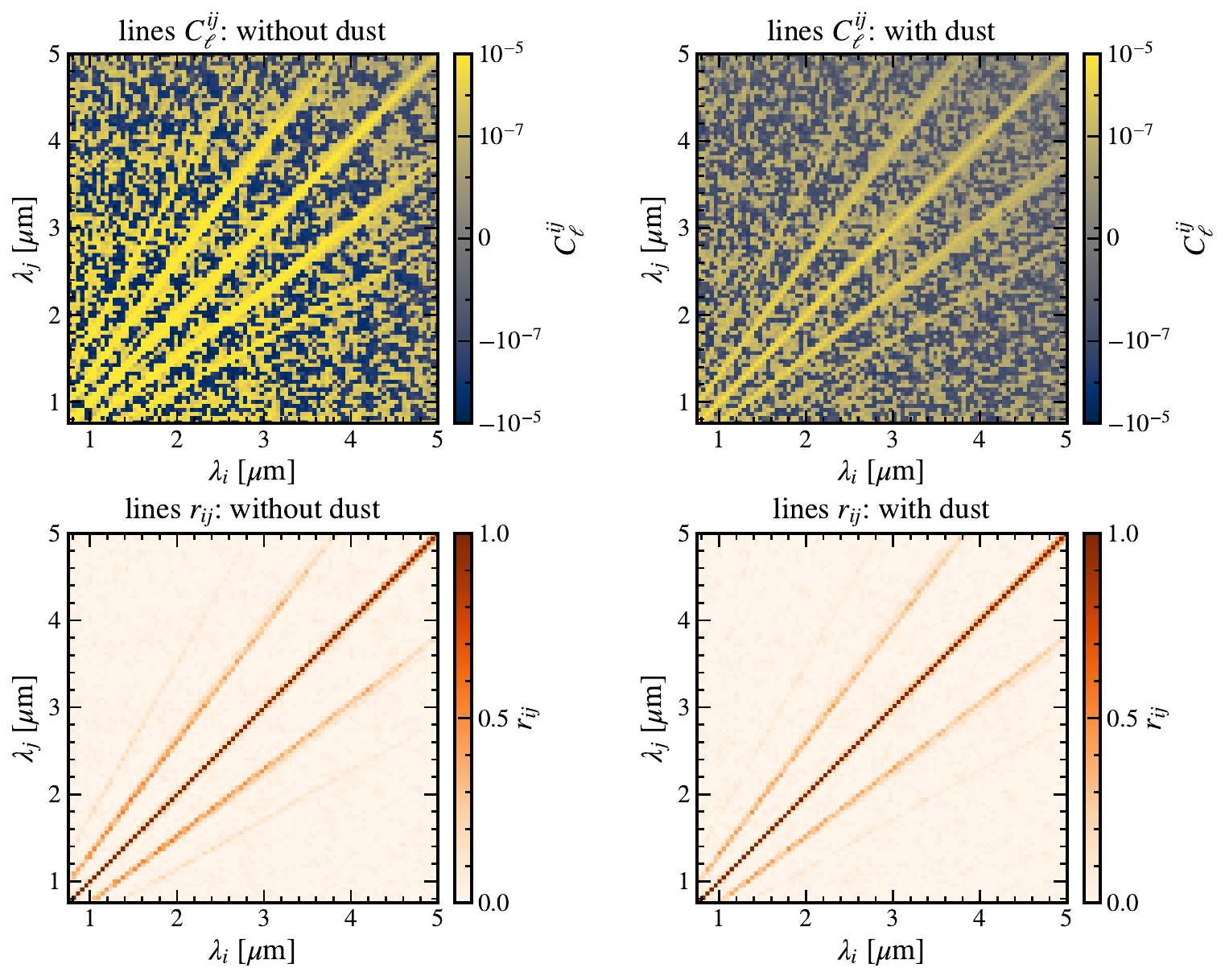}
    \caption{
Cross-channel angular power spectra and correlation matrices for the
SPHEREx-like observed-frame intensity maps at target multipole
$\ell=12000$, using the nearest saved bin $\ell=11571$, over the full
$0.75$--$5\,\mu$m band. Columns compare the no-dust case (left) and the
fiducial dust-attenuated case (right). Top: line-only cross-power for \ha, \hb, \oiii,
and \oii\ in units of $({\rm Jy\,sr^{-1}})^2\,{\rm sr}$, on a symmetric
logarithmic color scale so that positive and negative values are
visible; the same-redshift ridges of Equation~\ref{eq:ridge_results}
stand out against the noise-dominated off-ridge background, and dust
suppresses the overall amplitude while preserving the ridge geometry.
Bottom: the corresponding line-only normalized correlation matrix
$r_\ell^{ij}$; the ridge pattern is nearly unchanged between the two
columns, illustrating that uniform suppression largely cancels in
$r_\ell^{ij}$ and that dust information resides primarily in the
dimensional cross-power.
}
    \label{fig:dust-clij-rij}
\end{figure*}

To display the scale dependence of the signal, we show
\begin{equation}
    D_\ell^{ij}
    \equiv
    \frac{\ell(\ell+1)}{2\pi} C_\ell^{ij},
    \label{eq:dell}
\end{equation}
which gives a convenient view of the power per logarithmic interval in
angle.  We focus on the line-only maps in this first diagnostic
so that the intrinsic scale dependence of the LIM signal can be seen
before introducing continuum emission and cleaning.  We also restrict the
plot to $\ell\gtrsim1000$, where the finite-area mock has enough
independent modes for the spectra to be visually stable.

\begin{figure}[htbp]
\centering
\includegraphics[width=0.48\textwidth]{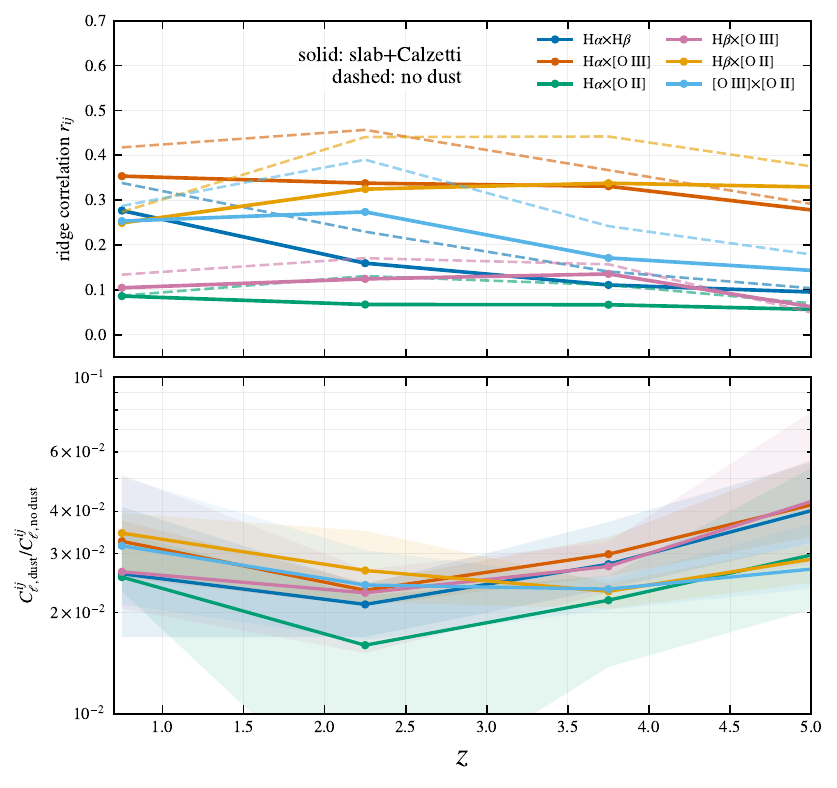}
\caption{
Redshift evolution of same redshift line-ridge observables for the
line-only maps at $\ell\simeq11571$.  Each curve follows the geometric locus where two
observed wavelength channels correspond to two different rest-frame
lines emitted at the same redshift.  Points show medians in
$\Delta z=0.5$ bins.  The top panel shows the normalized correlation
coefficient.
Solid curves include the fiducial SAM nebular dust attenuation, while
dashed curves show the no-dust case.  The bottom panel shows the ratio of the
dimensional ridge cross-power with dust to that without dust,
$C_{\ell,{\rm dust}}^{ij}/C_{\ell,{\rm no\,dust}}^{ij}$; shaded regions
show the 16th--84th percentile range within each redshift bin.  The
suppression is strongly line-pair and redshift dependent, demonstrating
that nebular dust cannot be absorbed into a single LIM amplitude
correction.
}
\label{fig:ridge_z}
\end{figure}

Figure~\ref{fig:cl_ell_spectra} summarizes the resulting scale
dependence of the line-only angular power spectra. The left panel shows
auto-spectra for representative observed channels. The spectra rise
toward high multipoles in the plotted quantity $D_\ell$, approaching at
$\ell\gtrsim5000$ the $D_\ell\propto\ell^2$ scaling expected when the
shot noise contribution from compact sources dominates. The
normalization varies by roughly an order of magnitude across channels,
with the bluest channel (0.94\,$\mu$m) carrying the most small-scale
power and the reddest (4.20\,$\mu$m) the least, because each channel
receives a different mixture of line emission from different redshifts
and rest-frame transitions. Thus, even before considering continuum
foregrounds, the LIM signal is strongly wavelength dependent.

The right panel of Figure~\ref{fig:cl_ell_spectra} shows a more directly tomographic diagnostic. We choose
an anchor channel at $\lambda_i=1.98\,\mu{\rm m}$, which corresponds to
\ha\ emission at $z\simeq2.02$. At this redshift, the associated
same-volume line wavelengths are approximately
$\lambda_j\simeq1.13\,\mu{\rm m}$ for \oii, $1.46\,\mu{\rm m}$ for \hb,
and $1.51\,\mu{\rm m}$ for \oiii. The solid curves in
Figure~\ref{fig:cl_ell_spectra} show that these same-redshift channels
maintain coherent positive cross-power with the \ha\ anchor across the
full multipole range, with amplitudes ordered as
\ha $>$ \oiii $>$ \hb $>$ \oii, whereas comparison channels not selected by
the line-ratio geometry are generally less stable and can fluctuate
through zero, as the 0.94 and 4.01\,$\mu$m curves illustrate. The
2.59\,$\mu$m comparison channel is the instructive exception: it
retains coherent positive cross-power because
$2.59/1.98\simeq\lambda_{\ha}/\lambda_{\oiii}$, so this channel pair
lies on the \ha$\times$\oiii\ locus at $z\simeq2.95$, sourced by the
\oiii\ emission entering the anchor channel. No observed channel is a
single redshift slice, and a channel pair can therefore intersect a
same-redshift locus for a line pair other than the one used to select
the anchor. This behavior is the $\ell$-space counterpart of the line
ridges seen in the cross-channel correlation matrices: the ridges are
not only localized in wavelength space, but also carry a characteristic
angular scale dependence.

This figure motivates the use of the full cross-channel matrix rather
than only auto-spectra. Auto-spectra measure the total line power in
each observed channel, but they do not by themselves identify which
redshift or rest-frame line dominates the signal. Cross-spectra between
channels selected by known rest-frame wavelength ratios provide a
cleaner tomographic handle, because they isolate emission from the same
comoving volume. In the following sections we therefore analyze both
the wavelength-space ridge structure and its response to dust
attenuation, continuum contamination, and spectral cleaning.

\subsection{Line-only Channel Correlations}
\label{sec:line_only_correlations}

We first isolate the line-only maps in order to examine the redshift
correlation pattern of the spectral cube. In this limit the continuum
is absent by construction, and the channel-correlation matrix is
dominated by correlations between rest-frame emission lines that enter
different observed-frame channels. The bottom row of
Figure~\ref{fig:dust-clij-rij} shows the line-only normalized
correlation matrix $r^{ij}_\ell$ of Equation~(\ref{eq:rij}) at
$\ell \simeq 11571$. The main diagonal corresponds to the
autocorrelation of each channel with itself and is therefore unity. The physically
interesting features are the off-diagonal ridges, which trace pairs of
rest-frame lines emitted at the same redshift.
For two lines with rest-frame wavelengths $\lambda_a$ and $\lambda_b$,
the same-redshift condition is
\begin{equation}
    \frac{\lambda_i}{\lambda_a}
    =
    \frac{\lambda_j}{\lambda_b}
    =
    1+z .
    \label{eq:ridge_results}
\end{equation}
Thus the slope of each ridge in the
$\lambda_i$--$\lambda_j$ plane is fixed by the ratio of rest-frame
wavelengths, while its visible extent is set by the observed wavelength
coverage and by the redshift range of the catalog. Ridges involving
[O~{\sc ii}] begin at higher redshift because [O~{\sc ii}] has the
shortest rest-frame wavelength among the four modeled transitions. Quantitatively, the median ridge correlations at $\ell\simeq11571$ are listed in Table~\ref{tab:ridges}. The H$\alpha\times$[O~{\sc iii}]
ridge is the strongest in the median, with $r\simeq0.30$, followed by
H$\beta\times$[O~{\sc ii}] with $r\simeq0.26$. The
H$\beta\times$[O~{\sc iii}] ridge is weaker, with median
$r\simeq0.05$. These differences are partly expected: the measured ridge amplitude depends on the line-luminosity normalizations, the redshift-dependent source population, dust attenuation, survey volume, and finite-area
mode sampling. The H$\beta\times$[O\,{\sc iii}] ridge is additionally penalized by geometry: the rest-wavelength ratio
$5007/4861 \simeq 1.03$ places this locus almost on the matrix
diagonal, so at the $R \sim 41$ spectral resolution of our channel
grid the ridge is only marginally separated from the channel
autocorrelation and is sampled by channel pairs whose spectral
responses partially overlap.

\begin{table}
\begin{center}
\begin{tabular}{lcccc}
\hline\hline
Line pair & $z_{\rm min}$ & $z_{\rm max}$ & $N_{\rm chan}$ & median $r$ \\
\hline
H$\alpha\times$H$\beta$                      & 0.54 & 6.62 & 86 & 0.120 \\
H$\alpha\times$[O~{\sc iii}]                 & 0.50 & 6.62 & 87 & 0.304 \\
H$\alpha\times$[O~{\sc ii}]                  & 1.01 & 6.62 & 80 & 0.061 \\
H$\beta\times$[O~{\sc iii}]                  & 0.54 & 8.99 & 88 & 0.053 \\
H$\beta\times$[O~{\sc ii}]                   & 1.01 & 9.29 & 87 & 0.263 \\
\mbox{[O~{\sc iii}]}$\times$\mbox{[O~{\sc ii}]} & 1.01 & 8.99 & 86 & 0.136 \\
\hline
\end{tabular}
\end{center}
\caption{
Line-ridge correlations at $\ell\simeq11571$ for the slab-geometry,
Calzetti-attenuated line-only maps. The redshift ranges give the analytic
overlap of the nominal $0.75$--$5~\mu{\rm m}$ observed-frame band for
each line pair. The channel count $N_{\rm chan}$ is the number of
discrete wavelength-channel pairs used to sample the corresponding
same-redshift ridge. The median $r$ values are measured along that ridge. \label{tab:ridges}
}
\end{table}

Figure~\ref{fig:ridge_z} summarizes how the same-redshift multi-line
signal evolves across the lightcone.  For each line pair, we identify
the wavelength-channel pairs satisfying
$\lambda_i=\lambda_{i,{\rm rest}}(1+z)$ and
$\lambda_j=\lambda_{j,{\rm rest}}(1+z)$, and then bin the resulting
ridge pixels in redshift.  The normalized ridge correlation varies strongly
among line pairs.  The strongest ridges are produced by line pairs such
as \ha$\times$\oiii\ and \hb$\times$\oii, while \ha$\times$\oii\ is
considerably weaker in this realization.  Dust changes the dimensional
cross-power even more dramatically: the dust/no-dust ratio is well below
unity and depends on both redshift and line pair.  This behavior shows
that dust attenuation cannot be represented as a single global amplitude
rescaling of the LIM signal.  Instead, it changes the relative weights
of different line pairs and redshift intervals.

\begin{figure*}[htbp]
\centering
\includegraphics[width=0.95\textwidth]{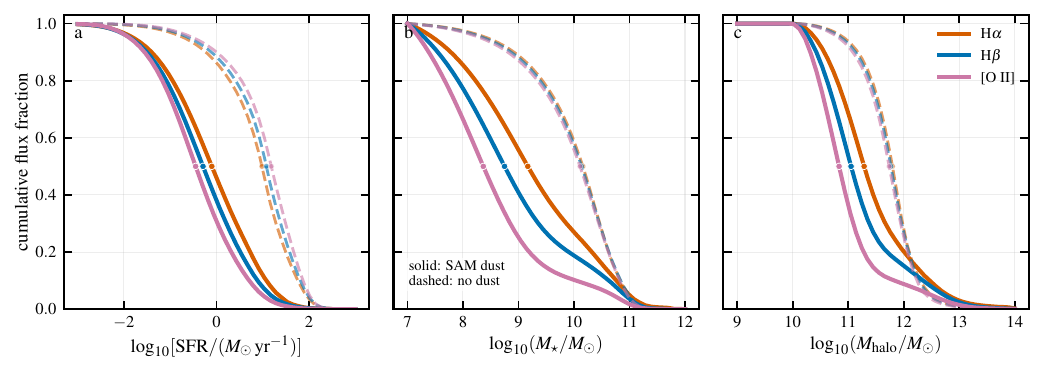}
\caption{
Cumulative source-level contribution to the observed integrated line
flux over the full lightcone.  The three panels rank galaxies by
star-formation rate, stellar mass, and halo mass, respectively.  The
quantity plotted is the fraction of the total observed line flux
produced by galaxies above the threshold on the horizontal axis.  Solid
curves include the fiducial SAM dust attenuation, while dashed curves
show the no-dust case.  Filled points mark the 50\% contribution
threshold for each curve.  Only galaxies whose observed line wavelength
lies inside the $0.75$--$5\,\mu{\rm m}$ band are included.  The no-dust
curves are similar because the intrinsic line luminosities are modeled
as fixed ratios relative to \ha.  Dust breaks this near-degeneracy by
introducing a wavelength- and galaxy-property-dependent attenuation,
shifting the dominant observed contribution away from the most actively
star-forming and most massive systems.
}
\label{fig:population_flux}
\end{figure*}

\subsection{Dust versus No-Dust}
\label{sec:dust_vs_nodust}

Before turning to the angular statistics, we first ask which galaxies
dominate the observed line budget. For each galaxy and each line, we
apply the same line-SFR calibration and dust model used in the map
construction, including a source only when its line falls in the
observed $0.75$--$5\,\mu$m band, and compute the cumulative
contribution to the integrated line flux as a function of galaxy
properties. Figure~\ref{fig:population_flux} shows the fraction of the
total observed flux contributed by galaxies above a given SFR, stellar
mass, or halo mass threshold. In the no-dust case, the curves for
different lines nearly coincide, because the intrinsic luminosities are
fixed ratios of \ha\ and the normalization cancels in the cumulative
fraction; half of the intrinsic flux comes from galaxies with
$\log_{10}({\rm SFR}/M_\odot\,{\rm yr}^{-1})\gtrsim 1.2$ and
$\log_{10}(M_\star/M_\odot)\gtrsim 10.1$. Dust shifts these thresholds
dramatically: in the attenuated maps the 50\% points move to
$\log_{10}({\rm SFR}/M_\odot\,{\rm yr}^{-1})\simeq -0.1$ and
$\log_{10}(M_\star/M_\odot)\simeq 9.2$ for \ha, more than an order of
magnitude in both quantities, with \oii\ shifted furthest because it is
attenuated most. The observed LIM flux therefore comes from a broader
and less massive population than the intrinsic emission would suggest,
a redistribution whose imprint on the cross-channel statistics we
quantify below.

We next compare the intrinsic and dust-attenuated channel-pair matrix.
Figure~\ref{fig:dust-clij-rij} shows line-only
cross-channel matrices at target multipole $\ell=12000$, using the nearest bin
$\ell=11571$. The left column shows the no-dust case and the right column
shows the fiducial dust-attenuated case. The top row shows the line-only cross-power, and the bottom row shows the line-only correlation matrix. For the line maps, dust affects both the absolute power and the relative
strengths of the same-redshift ridges. If a line has attenuation
$A_{\lambda,{\rm line}}$, its observed luminosity is
\begin{equation}
    L_{\rm line}^{\rm obs}
    =
    L_{\rm line}^{\rm int}
    \,10^{-0.4A_{\lambda,{\rm line}}}
    =
    L_{\rm line}^{\rm int}\,f_{\rm dust,line},
    \label{eq:line_dust_suppression}
\end{equation}
with $f_{\rm dust,line}$ the transmission factor of
Equation~\ref{eq:fdust_def}. Because the intensity map is linear in the
luminosities of its sources, the transmission enters the map as a
multiplicative weight, and the power spectrum, being quadratic in the
map, picks up two powers of it: an auto-spectrum is suppressed
approximately as $f_{\rm dust}^2$, and a cross-spectrum between lines
$X$ and $Y$ by the product $f_{{\rm dust},X}\,f_{{\rm dust},Y}$. The magnitude of this effect is set by the transmission factor shown in
Figure~\ref{fig:fdust_vs_av}: at a representative $A_V=1$, \ha\
transmits roughly $\sim18\%$ of its flux while \oii\ transmits only
$\sim5\%$. For a simple single-$A_V$ estimate, this implies that the
\ha$\times$\oiii\ ridge is suppressed by a factor
$f_{{\rm dust},\ha}\,f_{{\rm dust},\oiii}\simeq1.8\times10^{-2}$,
while the \hb$\times$\oii\ ridge is suppressed by
$f_{{\rm dust},\hb}\,f_{{\rm dust},\oii}\simeq4.5\times10^{-3}$. The quantity
$f_{{\rm dust},X,g}$ varies from galaxy to galaxy, so the measured
suppression is a luminosity-weighted population average rather than a
single number, and it differs between the clustering and shot noise terms.

\begin{figure*}[htbp]
    \centering   
    \includegraphics[width=\linewidth]{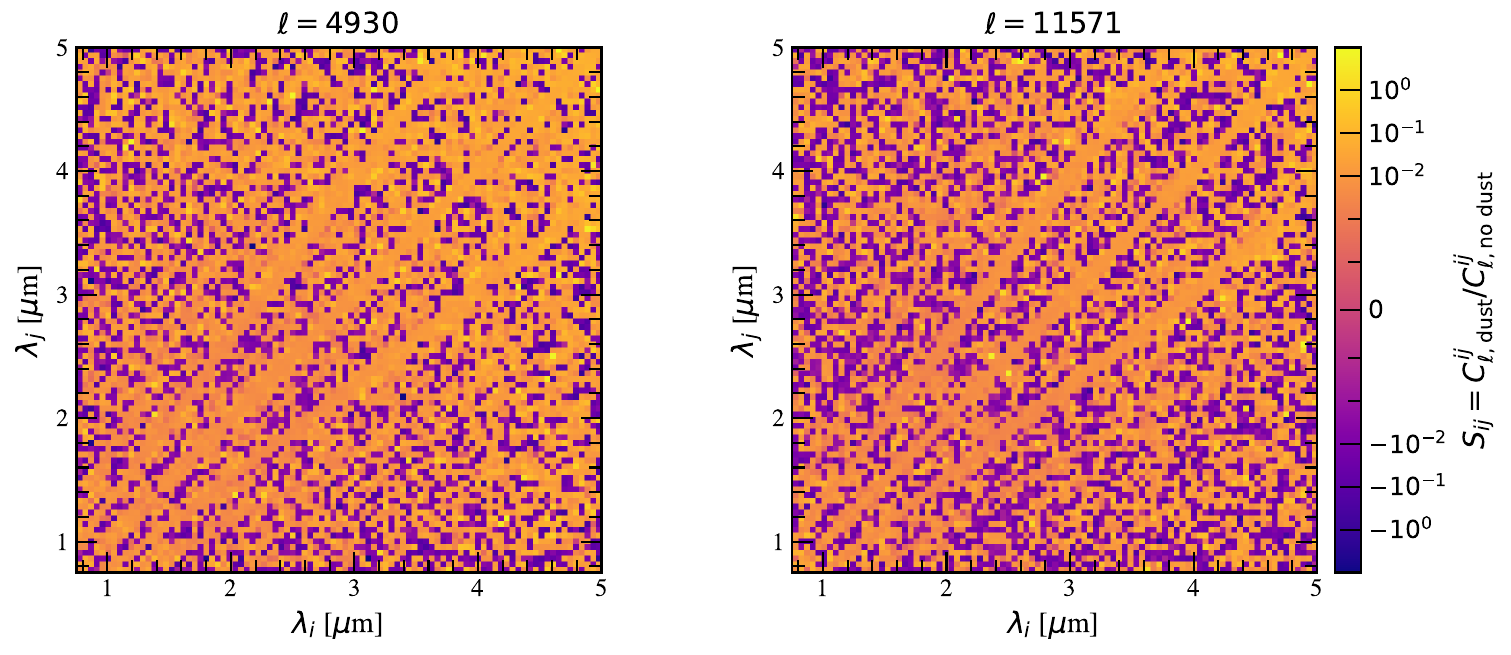}
    \caption{
    Dust-suppression matrix for the line-only LIM maps.  We show
    $S_{ij}(\ell)\equiv C_{\ell,\rm dust}^{ij}/C_{\ell,\rm no\ dust}^{ij}$
    as a function of observed wavelength channels $(\lambda_i,\lambda_j)$
    at $\ell=4930$ and $\ell=11571$.  The numerator and denominator are
    measured from the same SAM lightcone, with the only change being the
    inclusion or removal of nebular dust attenuation in the line luminosities.
    Values below unity indicate suppression of the line cross-power by dust.
    The color scale is symmetric logarithmic in order to display both the
    strongly suppressed regions and the small negative ratios that arise in
    weak, finite-area cross-spectrum estimates.  The structured pattern shows
    that dust does not act as a single global amplitude rescaling: its effect
    depends on observed wavelength, redshift, and line pair.
    }
    \label{fig:sij}
\end{figure*}

We compare the line-only map produced from the same SAM lightcone with and without nebular attenuation to reveal the map-level effect of dust. We then define the dust-suppression matrix as
\begin{equation}
    S_{ij}(\ell) \equiv
    \frac{C_{\ell,\rm dust}^{ij}}
         {C_{\ell,\rm no\ dust}^{ij}},
    \label{eq:sij}
\end{equation}
where $i$ and $j$ denote the observed wavelength channels. This ratio removes much of the geometric structure from the location of the line-ridges, thereby isolating the effect of dust on the cross-channel power spectrum. In Figure~\ref{fig:sij}, we show $S_{ij}$ at two
selected multipoles. The suppression varies systematically over the $(\lambda_i,\lambda_j)$ plane as a consequence of different observed wavelengths corresponding to different rest-frame lines and redshifts. Cross powers between shorter rest-frame wavelengths are more suppressed than the rest-frame \ha. Along the
same-redshift ridges, the suppression ranges from
$\sim3\times10^{-3}$ to $\sim3\times10^{-2}$ depending on redshifts and
line pair (Figure~\ref{fig:ridge_z}, lower panel), and is strongest
near $z\simeq2$, where actively dusty star-forming galaxies dominate
the line emission and the relative strengths of the ridges shift accordingly. As Table~\ref{tab:ridges} indicates, at $\ell\simeq11571$ the strongest medians in the dust-attenuated line-only maps are \ha$\times$\oiii\ ($r\simeq0.30$) and \hb$\times$\oii\ ($r\simeq0.26$), with \hb$\times$\oiii\ being down to $r\simeq0.05$. Dust is thus not just a single, scale-independent rescaling of the overall
$C_\ell^{ij}$ matrix, but a redshift- and line-pair-dependent restructuring
of it, as is immediately visible in the top panels of
Figure~\ref{fig:dust-clij-rij}. The normalized correlation matrix
$r_\ell^{ij}$, by contrast, remains largely unchanged (bottom row of
Figure~\ref{fig:dust-clij-rij}). This is not because dust affects all
line pairs equally, but because any suppression that acts multiplicatively
on individual channels cancels between the cross-power and the
auto-power normalization in Equation~(\ref{eq:rij}). The amplitude
information imprinted by dust therefore resides almost entirely in the
dimensional $C_\ell^{ij}$, while $r_\ell^{ij}$ isolates the geometric
ridge structure.

\subsection{Shot Noise and Clustering}
\label{sec:shot_noise}

The cross-channel power spectrum of the line maps contains two physically
distinct contributions: a clustering term from correlations between
different galaxies tracing the same large-scale density field, and a
shot noise term from the discreteness of the emitting galaxy population.
We can write it as 
\begin{equation}
C^{ij,{\rm total}}_\ell = C^{ij,{\rm clust}}_\ell
 + C^{ij,{\rm shot}}_\ell.
\label{eq:shot_clust_decomp}
\end{equation}
In an auto-channel
spectrum, shot noise is the Poisson contribution from individual
sources.  In a multi-line cross-channel matrix it also appears away from
the wavelength diagonal, because the same galaxy can emit more than one
line.  If two observed channels correspond to two rest-frame lines at the
same redshift, the same objects contribute flux to both channels and
generate an off-diagonal shot noise ridge.

\begin{figure*}[htbp]
    \centering
\includegraphics[width=\linewidth]{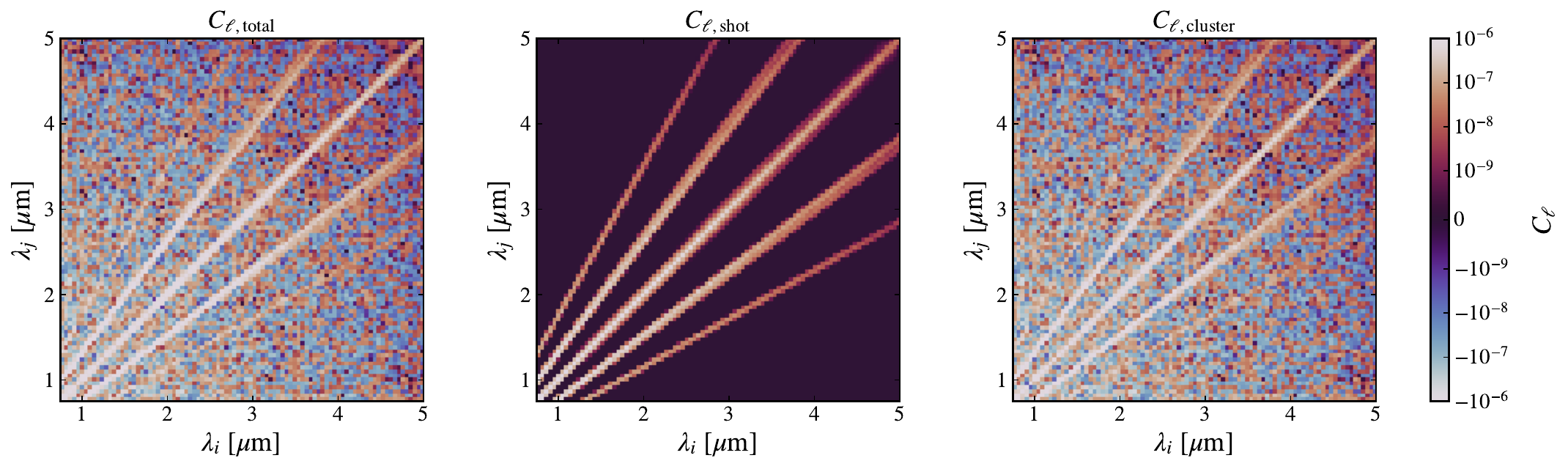}
    \caption{
    Shot noise decomposition of the line-only cross-channel power matrix at
    $\ell\simeq11571$.  From left to right we show the total line power,
    the estimated shot noise term, and the residual clustering contribution
    $C_{\ell,\rm cluster}^{ij}=C_{\ell,\rm total}^{ij}-C_{\ell,\rm shot}^{ij}$.
    The signed logarithmic color scale displays positive and negative matrix
    elements while preserving the line-ridge morphology.  The shot term is not
    limited to the wavelength diagonal: because the same galaxy can emit
    multiple lines, it also produces off-diagonal same-redshift ridges.  This
    demonstrates that the observed line-ridge amplitude contains both
    clustering and discreteness contributions.
    }
    \label{fig:shot}
\end{figure*}

For the line maps we estimate this contribution directly from the source
catalog by summing flux products over the survey footprint.  In the same unit of
surface brightness as Equation~(\ref{eq:cl}), the cross-channel shot
term is
\begin{equation}
  C_\ell^{ij,\rm shot}
  =
  \frac{1}{\Omega_{\rm survey}}
  \sum_g
  f_{\nu,i,g} f_{\nu,j,g},
  \label{eq:shot}
\end{equation}
where $f_{\nu,i,g}$ is the observed flux density from galaxy $g$ contributed
to channel $i$, including the line modeling and dust attenuation, and
$\Omega_{\rm survey}$ is the survey solid angle for the mock footprint.
When comparing to measured map power, we multiply by the same pixel window correction as used for the map level spectra. The residual clustering measure is then defined as: 
\begin{equation}
  C_\ell^{ij,\rm clust}
  =
  C_\ell^{ij,\rm total}
  -
  C_\ell^{ij,\rm shot}.
  \label{eq:cluster}
\end{equation}
 This subtraction should be interpreted as a diagnostic division of the mock map power, not a separate estimator for survey data where the shot noise would be modeled and fit simultaneously. Figure~\ref{fig:shot} shows the decomposition for the line-only maps at
$\ell\simeq 11571$.  The total matrix shows the expected same redshift
line ridges. The shot noise panel shows that shot noise is not only concentrated on the diagonal, but traces the line pairs because the same source of correlated emission contributes to different channels. Once this shared source term is subtracted, the residual clustering matrix shows much of the original structure but is reduced at the brightest line pairs. The median absolute value of total power in off-diagonal channels of the full channel matrix at this multipole is
$2.8\times10^{-8}\,({\rm Jy\,sr^{-1}})^2\,{\rm sr}$, and for the clustering contribution this is
$2.7\times10^{-8}\,({\rm Jy\,sr^{-1}})^2\,{\rm sr}$. The median shot
estimate is consistent with zero because most channel pairs do not share a
common same-galaxy line contribution.  However, the shot term is clearly
nonzero on selected same-redshift ridges and becomes increasingly important
for bright line pairs and smaller angular scales.  Thus, although the
full-matrix median at $\ell\simeq 11571$ is clustering dominated, the
interpretation of ridge amplitudes requires both clustering and shot noise
terms in the forward model.

\section{Continuum Cleaning}
\label{sec:cleaning}

The total observed maps are dominated by continuum emission, which exceeds the line signal by a few orders of magnitude. Unlike the
nebular lines, which appear as localized same-redshift ridges in the
cross-channel correlation matrix, the stellar continuum is spectrally
smooth and therefore produces broad correlations across wavelength. Before cleaning,
$r_{ij}$ is close to unity over large regions of wavelength space of the continuum emission, and
the line ridge structure is largely hidden.  We therefore apply spectral
continuum cleaning to the total map cube before measuring line
cross-correlations.

Our baseline cleaning method is a PCA-based method along
the spectral direction.  We write the map cube as a matrix
$X_{p i}$, where $p$ labels sky pixels and $i$ labels wavelength
channels.  PCA treats the sky pixels as samples and the wavelength
channels as features.  Concretely, after subtracting the mean spectrum,
we form the $N_\lambda \times N_\lambda$ channel covariance matrix
$\Sigma_{ij} = \sum_p X_{pi} X_{pj} / N_{\rm pix}$, which measures how
strongly the intensity fluctuations in each pair of channels vary
together across the sky.  Diagonalizing $\Sigma$ yields a set of
orthonormal eigenvectors $v_k$, each of which is a template spectrum
across the $N_\lambda$ channels, ordered by their eigenvalues, i.e., by
the amount of map variance each template accounts for.  Any pixel
spectrum can be expressed as a sum over these templates, so removing the
first $N_{\rm PCA}$ of them amounts to subtracting, pixel by pixel, the
best-fit combination of the most dominant spectral shapes in the data.
The leading principal components describe the spectral
modes with the largest variance across the map.
Because the continuum is bright and spectrally coherent, these leading
modes are dominated by smooth continuum structure.  Removing them
suppresses the continuum while leaving behind residual fluctuations that
contain the line signal, noise, and any continuum component not captured
by the removed modes.

For a choice of $N_{\rm PCA}$ removed modes, we compute the first
$N_{\rm PCA}$ spectral principal components of the total data cube and
collect them in the matrix
\begin{equation}
    V_N = \left(v_1, v_2, \ldots, v_{N_{\rm PCA}}\right),
\end{equation}
where each column vector $v_k$ is a spectral eigenvector across the
wavelength channels. If the cube has $N_\lambda$ wavelength channels,
then $V_N$ has dimensions $N_\lambda \times N_{\rm PCA}$. The cleaned
cube is then
\begin{equation}
    X_{\rm clean}
    =
    X - X V_N V_N^{T},
\end{equation}
where $X$ is written as a two-dimensional matrix with sky pixels as rows
and wavelength channels as columns. Thus, $X V_N V_N^T$ is the projection
of the data onto the subspace spanned by the first $N_{\rm PCA}$ spectral
modes, and subtracting it removes the smooth spectral modes identified
by PCA. In practice this operation is applied to the same total cube used in the
correlation analysis.  The procedure is intentionally blind as the PCA
modes are learned from the observed total cube, not from the line-only
or continuum-only truth cubes.  The truth cubes are used only to evaluate
the performance of the cleaning at the end stage.

A central complication is that PCA cleaning is not a perfect continuum
subtraction.  The PCA modes are determined from the total cube, and
therefore any component of the line signal that resembles the removed
spectral modes will also be partially subtracted.  The cleaned residual
is consequently a filtered version of the true line cube, not an
unbiased estimate of it.  We quantify this filtering with a line transfer
function.

To measure the transfer function, we inject a small copy of the line-only signal cube
into the total cube and pass both cubes through exactly the same cleaning
procedure.  If ${\cal C}$ denotes the PCA cleaning operator, we define
the cleaned response to the injected line signal as
\begin{equation}
  \Delta_{\rm clean}
  =
  \frac{
  {\cal C}\!\left(I_{\rm total}+a I_{\rm line}\right)
  -
  {\cal C}\!\left(I_{\rm total}\right)}
  {a},
  \label{eq:transfer_delta}
\end{equation}
where \(a\) is the injection amplitude.  This finite difference isolates
the part of the input line cube that survives the cleaning.  If the
cleaning preserved the line signal perfectly, then
\(\Delta_{\rm clean}=I_{\rm line}\).  If the cleaning removed part of the
line signal, then \(\Delta_{\rm clean}\) has reduced amplitude.

For each wavelength channel, we define the same channel line transfer as
\begin{equation}
  T_i(\ell)
  =
  \frac{
  C_\ell\!\left[\Delta_{{\rm clean},i}, I_{{\rm line},i}\right]
  }
  {
  C_\ell\!\left[I_{{\rm line},i}, I_{{\rm line},i}\right]
  } .
  \label{eq:transfer}
\end{equation}
This quantity measures the fraction of the line auto power in channel
\(i\) that is retained after cleaning.  Values \(T_i(\ell)\simeq1\)
indicate nearly lossless recovery of the line mode, while
\(T_i(\ell)<1\) indicates that PCA has projected out part of the line
signal.  Very small or negative values indicate that the cleaned map is
not a reliable tracer of the original line field in that channel and
multipole bin.

To compare different choices of \(N_{\rm PCA}\), we compress the
channel-dependent transfer function \(T_i(\ell)\) into a scalar summary
by taking its median over wavelength channels,
\begin{equation}
    T_{\rm med}(\ell)
    =
    {\rm median}_i\, T_i(\ell).
\end{equation}
This quantity is used only as a diagnostic of the typical line
attenuation caused by cleaning; the full transfer function remains
channel dependent.  We also quote the corresponding median line loss,
\begin{equation}
    f_{\rm loss}(\ell)
    =
    1 - T_{\rm med}(\ell).
\end{equation}

\begin{figure}[htbp]
    \centering    \includegraphics[width=\linewidth]{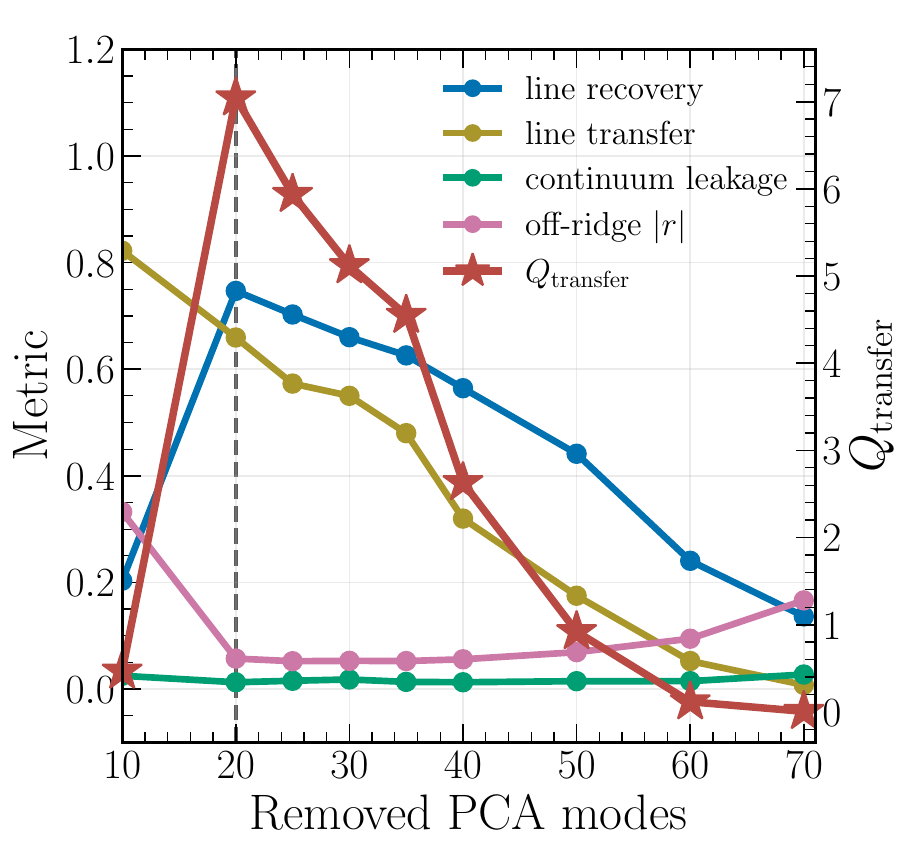}
    \caption{
    PCA-cleaning performance as a function of the number of spectral modes removed, evaluated near the target multipole $\ell=12000$ using the nearest available bin, $\ell=11571$. The red star symbols show the combined transfer quality metric, $Q_{\rm transfer}$, plotted on the right-hand axis. The dashed vertical line marks the preferred choice, PCA20, which gives the best balance between recovering the line signal and suppressing continuum leakage.
    }
    \label{fig:pca-vs-modes}
\end{figure}

\begin{figure}[htbp]
    \centering
\includegraphics[width=\linewidth]{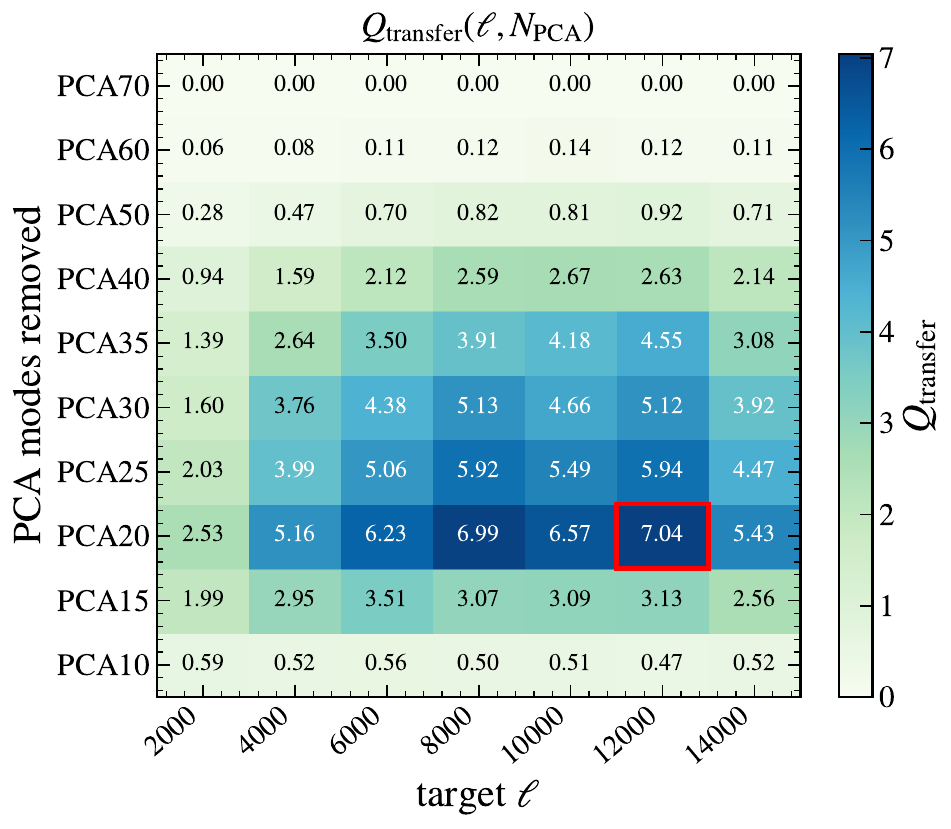}
    \caption{
    Transfer-quality metric, $Q_{\rm transfer}(\ell, N_{\rm PCA})$, evaluated
    over a grid of target multipoles and numbers of PCA modes removed. Each
    cell gives the value of $Q_{\rm transfer}$ for the corresponding cleaning
    choice. The red box marks the best-performing configuration in this grid,
    PCA20 at target $\ell=12000$. The metric is largest for intermediate cleaning strength: removing too few modes leaves continuum residuals, while removing too many modes suppresses the line signal itself. This motivates the use of PCA15--PCA25 as the fiducial range for the continuum cleaning.
    }
    \label{fig:pca-transfer-matrix}
\end{figure}

We assess each $N_{\rm PCA}$ on the basis of four complementary
diagnostics at $\ell \simeq 11571$: the line-recovery correlation
$Q_{\rm line}$, median line transfer $T_{\rm med}$, residual continuum
leakage $Q_{\rm cont}$, and off-ridge residual correlation $Q_{\rm off}$. These diagnostics balance two
conflicting requirements for cleaned cubes: they should contain the line signal but reject the smooth foreground continuum. The first diagnostic, $Q_{\rm line}$, measures the extent to which the cleaned residual map traces the true line-only truth map:
\begin{equation}
    Q_{\rm line}
    =
    {\rm median}_i\,
    r_\ell\!\left(I^{\rm clean}_i,I^{\rm line}_i\right),
    \label{eq:qline}
\end{equation}
where $I^{\rm clean}_i$ is the cleaned residual in channel $i$ and
$I^{\rm line}_i$ is the line-only truth map in channel $i$. Higher values mean that spatial structure in the cleaned residual agrees with the true LIM structure. This diagnostic is sensitive to overly aggressive PCA cleaning, which can remove desired line signal.

The second diagnostic, $Q_{\rm cont}$, measures how well continuum
leakage is suppressed, and we express it as
\begin{equation}
    Q_{\rm cont}
    =
    {\rm median}_i\,
    \left|r_\ell\!\left(I^{\rm clean}_i,I^{\rm cont}_i\right)\right|,
    \label{eq:qcont}
\end{equation}
where $I^{\rm cont}_i$ is the continuum-only truth map in channel $i$. This metric should be small, indicating weak correlations between the cleaned residual and the continuum truth map. This statistic penalizes under cleaning,
where too few PCA modes are removed and smooth continuum structure
remains in the residual cube.

The third statistic is the median line transfer, $T_{\rm med}$, which is calculated from the median channel-dependent transfer function $T_i(\ell)$. This measures the typical amplitude response of the line
signal, i.e. the median fraction of line power remaining after cleaning. This is important because a cleaned map may
remain highly correlated with the true line field while still having a suppressed amplitude.  Thus \(Q_{\rm line}\) measures whether the
surviving residual has the correct spatial pattern, whereas
\(T_{\rm med}\) measures how much of the line amplitude is retained.

The fourth statistic, $Q_{\rm off}$, measures spurious correlations in the residual map. We express it as
\begin{equation}
    Q_{\rm off}
    =
    {\rm median}_{(i,j)\notin{\cal R}}
    \left|r^{\rm clean}_{ij}\right|,
    \label{eq:qoff}
\end{equation}
where ${\cal R}$ is the set of channel pairs expected to contain the 
same redshift line-ridges. The line signal produces correlations along specific ridges in the wavelength--wavelength plane, set by the ratios of the rest-frame line wavelengths. Broad correlations away from these
ridges are therefore not part of the desired multi-line signal. A large $Q_{\rm off}$ indicates coherent residual structure, usually from imperfect continuum subtraction, which could mimic or obscure the ridge
features. We quantify the performance of PCA cleaning by comparing these four quantities as a function of the number of removed spectral modes.

Figure~\ref{fig:pca-vs-modes} illustrates the behavior of these four
diagnostics as a function of the number of PCA modes removed. The useful
cleaning regime is not the one that simply minimizes the
residual variance, but the one that preserves the line field while
suppressing continuum leakage. The blue curve measures the line recovery correlation, \(Q_{\rm line}\),
i.e. whether the cleaned maps trace the spatial fluctuations of the
line-only truth. This quantity rises steeply up to PCA20, where it peaks, and declines gradually for more aggressive cleaning mechanisms. The decrease at large \(N_{\rm PCA}\) shows that the PCA
projection is no longer removing only smooth continuum modes but it is also
removing desired line signals. The gold curve shows the median line transfer, \(T_{\rm med}\), which
measures the typical amplitude response of the line signal.  Unlike the line recovery correlation, the transfer function almost monotonically decreases
as more PCA modes are removed.  This means that a
cleaned map can remain spatially correlated with the true line field
while it could still have its line amplitude suppressed. Thus
\(Q_{\rm line}\) and \(T_{\rm med}\) are complementary to each other because
\(Q_{\rm line}\) tests whether the right spatial pattern survives after cleaning, while
\(T_{\rm med}\) estimates how much of the line amplitude survives. The continuum leakage, $Q_{\rm cont}$ (green curve), remains consistently low over the range of cleaning levels shown. The off-ridge correlation $Q_{\rm off}$ is minimized around PCA25 and begins to increase for very aggressive cleaning due to noise contamination of the residual, but the general trend is that the leading modes have significant continuum contamination, and these are removed in PCA20 to PCA25, thus lowering $Q_{\rm off}$.

We combine these diagnostics into a single metric, the transfer quality
\begin{equation}
    Q_{\rm transfer}
    =
    \frac{Q_{\rm line}\,T_{\rm med}}
    {Q_{\rm cont}+Q_{\rm off}} .
\end{equation}
This scalar diagnostic rewards line recovery and amplitude preservation,
while penalizing both continuum leakage and unwanted residual structure. It serves as a useful proxy to evaluate the effectiveness of PCA
cleaning.  In this realization, the maximum occurs at PCA20, which gives
$Q_{\rm line}=0.747$, $Q_{\rm cont}=0.0127$, $T_{\rm med}=0.659$, and
$Q_{\rm off}=0.0572$, corresponding to $Q_{\rm transfer}=7.04$. PCA25 is
a nearby, more aggressive choice, with $Q_{\rm line}=0.703$,
$Q_{\rm cont}=0.0156$, $T_{\rm med}=0.573$, $Q_{\rm off}=0.0523$, and
$Q_{\rm transfer}=5.94$. We therefore adopt PCA20 as the fiducial,
metric-optimized cleaner and treat PCA25 as a robustness case with
stronger cleaning.

Figure~\ref{fig:pca-transfer-matrix} extends this comparison to a 2D grid in a target multipole $\ell$, and the number of PCA modes removed $N_{\rm PCA}$. In each cell, we show the combined diagnostic \(Q_{\rm transfer}\), which rewards line recovery and line transfer, and penalizes continuum leakage and off-ridge residual correlations.   The purpose of this plot is to test whether the preferred cleaning choice is
a special feature of one multipole bin, or whether it remains stable
over a broader range of angular scales.  The main result is that it is
not an isolated feature: intermediate cleaning strengths, PCA15--PCA25,
give consistently high performance over
\(4000\lesssim\ell\lesssim14000\), with PCA20 providing the largest \(Q_{\rm transfer}\) for every multipole column in the grid. Highly aggressive cleaning, \(N_{\rm PCA}\gtrsim50\), yields poor results at all \(\ell\), because the PCA projection removes too much of the line signal. Conversely, the mildest cleaning with PCA10 leaves the continuum residuals dominant over the off-ridge statistics. 

The best individual cell in this grid is PCA20 at
\(\ell\simeq11571\). However, the figure shows that this is not
optimal in isolation. Nearby choices, such as PCA20 at
\(\ell=8000\)--10000 and PCA25 at \(\ell=8000\)--12000, also give high values of the diagnostic. This grid confirms PCA20 as our default, metric-optimized choice, while
PCA25 provides a more aggressive robustness test that removes additional
modes, lowers residual correlations, and sacrifices some line transfer. The comparison across \(\ell\) provides a useful robustness check that the adopted cleaning
prescription is not driven by one particular Fourier bin.

In the mocks we use truth-level quantities such as $Q_{\rm transfer}$ and
$Q_{\rm line}$ to identify the cleaning regime, but these diagnostics are
not available for real data, where the true line-only and continuum-only
fields are unknown and quantities such as the line recovery and continuum
leakage cannot be measured directly. Two data-driven diagnostics take
their place. The first is the behavior of the cleaned
wavelength--wavelength correlation matrix as a function of $N_{\rm PCA}$:
one can monitor the median off-ridge correlation,
\begin{equation}
    Q_{\rm off}^{\rm data}(N_{\rm PCA})
    =
    {\rm median}_{(i,j)\notin{\cal R}}
    \left|r_{\ell,{\rm clean}}^{ij}\right|,
\end{equation}
where ${\cal R}$ is the set of geometrically predicted same-redshift
line ridges. This statistic uses only the observed data and tests whether
broad continuum-like correlations remain away from the expected line
features. The second is an injection-based estimate of the line transfer:
we treat the total map as the observed data, inject a known line-like
signal, rerun the identical PCA cleaning procedure, and measure how much
of the injected signal is recovered. A useful cleaning range is indicated
when $Q_{\rm off}^{\rm data}$ has reached a plateau while the injected-line
recovery remains acceptably high, so that further increases in
$N_{\rm PCA}$ remove line signal without appreciably reducing the
residuals.

\begin{figure}[htbp]
    \centering
    \includegraphics[width=0.48\textwidth]{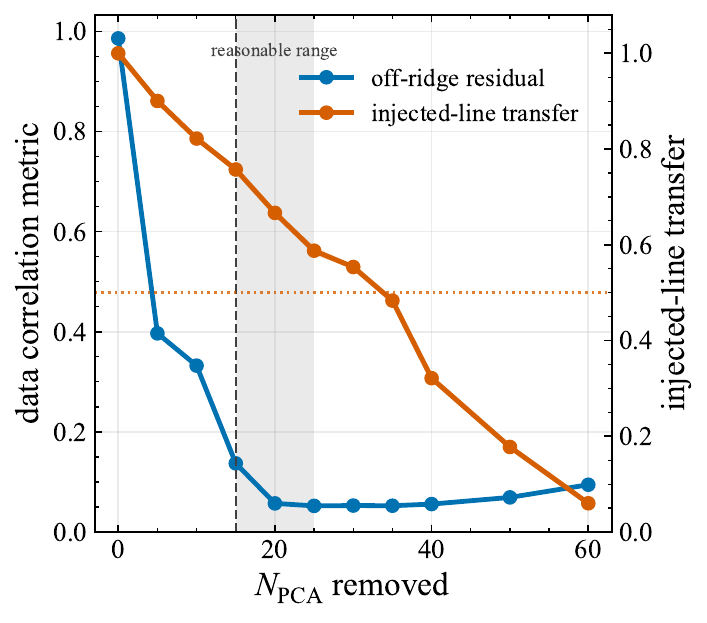}
    \caption{
    Real-data-like PCA stopping test at $\ell\simeq11571$.  The total map is
    treated as the observed data, and a known line-only template is injected
    into this map before rerunning the same PCA cleaning procedure.  The blue
    curve shows the median off-ridge correlation in the cleaned
    wavelength--wavelength matrix, which traces broad continuum-like residuals.
    The orange curve shows the median recovered transfer of the injected line
    signal, and the dotted orange line marks the minimum transfer threshold used
    here.  The shaded band marks the range $15\leq N_{\rm PCA}\leq25$, where
    the off-ridge residuals have largely plateaued while the injected line
    response remains substantial.  This range therefore gives a practical
    cleaning choice without using continuum or line truth maps directly.
    }
    \label{fig:pca-injection-stop}
\end{figure}

The plot shown in Figure~\ref{fig:pca-injection-stop} illustrates this test for a chosen \(\ell\) and various levels of cleaning (characterized by $N_{\rm PCA}$, the number of removed PCA modes). In the case of mild cleaning (low $N_{\rm PCA}$), a large fraction of the injected signal is recovered, but the wavelength-wavelength correlation matrix remains significantly populated in off-ridge regions, suggesting persistent continuum residuals. As $N_{\rm PCA}$ is increased, these off-ridge residuals decrease, but so does the recovery of the injected signal. A reasonable cleaning level is reached when $Q_{\rm off}^{\rm data}(N_{\rm PCA})$ reaches a plateau and remains relatively flat, while the injection recovery remains acceptably high. For this realization, the injection test identifies
$N_{\rm PCA}\simeq15$--25 as a reasonable range. Very aggressive cleaning ($N_{\rm PCA}\gtrsim35$) continues to remove signal but provides no significant reduction in the residual matrix. This shows that we can define a robust and practically applicable criterion for setting the number of removed PCA modes. In this way, one can find a balance between reducing observed residuals and calibrating line recovery using an injection test. 

\section{Learning from the Mock Data}
\label{sec:inference}

The forward model we developed gives us a controlled setting in which
the true line and continuum components are known separately. We use this mock map to verify whether multi-line
LIM observables can recover astrophysical parameters in the presence of a dominant continuum component after the continuum cleaning. The mock analysis lets us separate three questions that are entangled in
real data.  First, we assess the identifiability of the line amplitude ($A_{\rm line}$) and dust attenuation ($A_{\rm dust}$) through their respective signatures in the cross-channel spectra.  Second, we can quantify the effect of continuum
cleaning, which suppresses smooth foregrounds but also attenuates part of the line signal. Third, we determine which portions of the wavelength--wavelength matrix are most informative, specifically whether same redshift line-ridges alone are sufficient, or whether additional information from off-ridge correlations and the entire matrix offers improvement. The inferred parameter values can thus reveal which parts of the LIM data vector serve as sensitive probes of the galaxy population, versus those that are dominated by cleaning residuals or transfer effects.  

Successful inference of $A_{\rm line}$ would imply that the clean line-ridge structure retains information about total star-formation-driven line emission, while robust inference of $A_{\rm dust}$ would indicate that relative line-ridge intensities reflect nebular attenuation. On the other hand, the presence of broad degeneracies or biased parameter estimates would highlight limitations in extracting information from cleaning residuals, template correlations, and transfer. At fixed multipole $\ell$, the data vector is the cross-channel angular power spectrum matrix,
\begin{equation}
    d_{ij}(\ell) \equiv C_\ell(\lambda_i,\lambda_j),
\end{equation}
or equivalently the subset of matrix elements selected for the
likelihood.  

For the raw total maps, our model incorporates both the line and continuum contributions:
\begin{equation}
    C_{\ell,ij}^{\rm model}
    =
    A_{\rm line}\,
    C_{\ell,ij}^{\rm line}(A_{\rm dust})
    +
    A_{\rm cont}\,
    W_{ij}(\alpha_{\rm cont})\,
    C_{\ell,ij}^{\rm cont}.
    \label{eq:inference_model}
\end{equation}

The first term represents the nebular line signal.  The parameter
\(A_{\rm line}\) rescales the overall line amplitude, absorbing
uncertainties in the line-luminosity relation, bias, and
global intensity normalization. The parameter \(A_{\rm dust}\) is
the effective nebular attenuation of the line template:
\(A_{\rm dust}=0\) corresponds to the no-dust line model, while
\(A_{\rm dust}=1\) corresponds to the fiducial SAM dust model. The second term represents continuum power.  The parameter
\(A_{\rm cont}\) rescales the continuum-template amplitude, while
\(\alpha_{\rm cont}\) allows a smooth wavelength-dependent tilt through
\(W_{ij}(\alpha_{\rm cont})\). This tilt characterizes smooth residual continuum, or calibration, misalignments rather than line-like feature specific mismatches. The line component is therefore allowed to have an amplitude and dust attenuation. The continuum component is given only a limited amount of smooth spectral freedom.

Such an expression is derived from the usual method to describe smooth foreground contaminants as low-dimensional spectral contaminations. 
The foreground, like continuum, residuals, unlike the line signal with correlations lying along narrow same-redshift ridges, change slowly with the frequency or wavelength. The 21 cm and LIM data analyses frequently employ smooth spectral foreground models where the data are projected or modeled as low-order polynomials of frequency, for instance \citep[see, e.g.,][]{wang2006-21cm, LiuTegmark2011, Switzer2019}. This type of parameterization imposes a slow spectral tilt onto our continuum templates directly.  For example, if the amplitude of the remaining continuum in channel \(i\) varies with frequency in the way \((\lambda_i/\lambda_{\rm piv})^{\alpha_{\rm cont}/2}\) this implies that the cross power spectrum of the channels \(i\) and \(j\) receives a modulation by the factor:

\begin{equation}
    W_{ij}(\alpha_{\rm cont})
    =
    \left[
    \left(\frac{\lambda_i}{\lambda_{\rm piv}}\right)^{\alpha_{\rm cont}}
    \left(\frac{\lambda_j}{\lambda_{\rm piv}}\right)^{\alpha_{\rm cont}}
    \right]^{1/2}.
\end{equation}
This product is the normalization which is accounted for by the choice of a reference wavelength \(\lambda_{\rm piv}\), meaning that \(\alpha_{\rm cont}\) affects only the broad spectrum of the tilt and not its normalization (which is controlled by \(A_{\rm cont}\)). Hence, this term is not aiming to represent the physical contribution from stellar continuum, but rather serving as a parameter accounting for remaining smooth continuum misalignments which result from data calibration or masks.

The dust parameter $A_{\rm dust}$ is an effective LIM dust parameter.  It
morphs between two line templates:
\begin{equation}
    C_{\ell}^{\rm line}(A_{\rm dust}=0)
    =
    C_{\ell}^{\rm line,no\ dust},
\end{equation}

\begin{equation}
    C_{\ell}^{\rm line}(A_{\rm dust}=1)
    =
    C_{\ell}^{\rm line,fiducial\ dust}.
\end{equation}
Thus $A_{\rm dust}=0$ corresponds to the no-dust line cube, while
$A_{\rm dust}=1$ corresponds to the fiducial SAM nebular-dust model used
throughout this work.  Values between these limits interpolate the
strength of dust suppression in the line template.  This parameter should
not be interpreted as the dust attenuation of an individual galaxy; it is
a population-level effective parameter inferred from the LIM
cross-channel matrix.

For cleaned maps, we apply the same likelihood either to the cleaned
matrix or to the transfer-corrected cleaned matrix.  In the most
optimistic case, continuum cleaning removes the smooth continuum
subspace well enough that the model can be reduced to the line term,
\begin{equation}
    C_{\ell,ij}^{\rm model}
    =
    A_{\rm line}\,
    C_{\ell,ij}^{\rm line}(A_{\rm dust}).
\end{equation}
In practice, we also test versions with a residual continuum amplitude
to quantify how much continuum information remains after cleaning.  The
comparison between raw and cleaned inference therefore measures both the
benefit of continuum removal and the cost of line transfer loss.

We assume a Gaussian likelihood for the selected matrix elements,
\begin{equation}
    -2\ln{\cal L}
    =
    \sum_{(i,j)\in{\cal M}}
    \frac{
    \left[
    C_{\ell,ij}^{\rm data}
    -
    C_{\ell,ij}^{\rm model}(\boldsymbol{\theta})
    \right]^2
    }{\sigma_{ij}^2}
    + {\rm const.},
    \label{eq:matrix_likelihood}
\end{equation}
where ${\cal M}$ is the likelihood mask and
$\boldsymbol{\theta}=\{A_{\rm line},A_{\rm dust},A_{\rm cont},
\alpha_{\rm cont}\}$ is the parameter vector.  The mask can be chosen to be one of several: the full matrix, the expected line-ridge regions, or even a subset focused on dust-sensitive modes like the Balmer-ridge cells.  The uncertainties
$\sigma_{ij}$ are assigned from the mock covariance model used in the
inference test.  In later survey applications these should be replaced by
a covariance including instrumental noise, sample variance, masking, and
cleaning-induced mode coupling. In the recovery tests below we deliberately inject an off-fiducial
truth value, $A_{\rm dust} = 0.8$. This necessitates a model that can discriminate between moderate values of dust attenuation, not merely identify whether there is no dust or pure dust. 

\begin{deluxetable}{lccc}
\tablecaption{Mock parameter recovery before and after PCA cleaning. \label{tab:prepost_inference}}
\tablehead{
\colhead{Stage} & \colhead{Parameter} & \colhead{Truth} & \colhead{Recovered value}
}
\startdata
\multicolumn{4}{c}{Pre-cleaning continuum fit} \\
Raw full matrix & $A_{\rm cont}$      & 1.00 & $1.00^{+0.002}_{-0.001}$ \\
Raw full matrix & $\alpha_{\rm cont}$ & 0.50 & $0.50^{+0.004}_{-0.004}$ \\
\hline
\multicolumn{4}{c}{Post-cleaning PCA20 line fit} \\
PCA20 ridge & $A_{\rm line}$ & 1.00 & $1.00^{+0.08}_{-0.07}$ \\
PCA20 ridge & $A_{\rm dust}$ & 0.80 & $0.80^{+0.01}_{-0.01}$ \\
PCA20 ridge & $A_{\rm cont}$ & 1.00 & $1.00^{+0.03}_{-0.03}$ \\
PCA20 full  & $A_{\rm line}$ & 1.00 & $1.00^{+0.05}_{-0.06}$ \\
PCA20 full  & $A_{\rm dust}$ & 0.80 & $0.80^{+0.01}_{-0.01}$ \\
PCA20 full  & $A_{\rm cont}$ & 1.00 & $1.00^{+0.02}_{-0.02}$ \\
\enddata
\end{deluxetable}

We sample the posterior with an \textit{emcee} \citep{emcee} sampler based on affine-invariant MCMC. All four parameters, $\{A_{\rm line}, A_{\rm dust}, A_{\rm cont}, \alpha_{\rm cont}\}$, are sampled directly within their prior ranges, and so the posterior is of the form 
\[ P(\boldsymbol{\theta}|d) \propto {\cal L}(d|\boldsymbol{\theta})P(\boldsymbol{\theta}) \] 
with relatively broad top-hat priors, selected to be wider than the injected mock truth. For each inference, we calculate the marginalized posterior constraints and the correlations between parameters, as well as a posterior predictive model. 

Our mock inference tests a set of issues: we verify the likelihood can accurately recover the injected amplitude $A_{\rm line}$; we demonstrate it can recover $A_{\rm dust}$, i.e., that it can distinguish from no-dust or mild dust; we quantify degeneracies between the line amplitude, dust parameters and residual continuum. We demonstrate the effect of the PCA-cleaned matrix on these degeneracies by comparing the raw Total and PCA-cleaned maps, and in doing so, illustrate how the overwhelming continuum matrix can dominate the likelihood, forcing much of its freedom into fitting smooth broadband emission.

\begin{figure}[t]
    \centering
    \includegraphics[width=\linewidth]{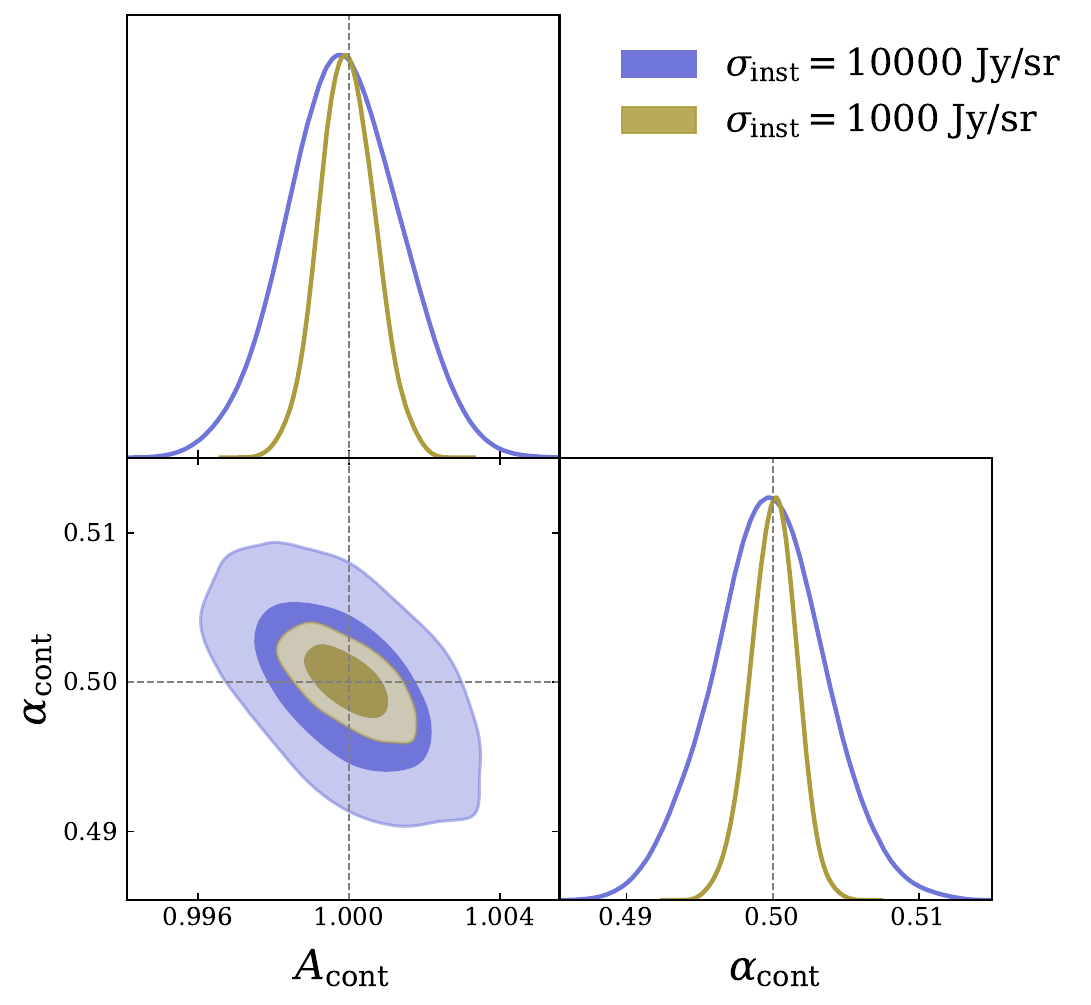}
    \caption{
Pre-cleaning continuum-parameter inference from the wavelength cross-power
matrix. The contours show the posterior constraints on the two continuum
model parameters, $A_{\rm cont}$ and $\alpha_{\rm cont}$. The correlation
between parameters reflects the fact that changes in the continuum amplitude
and broad spectral tilt both affect the broadband continuum power.
}
\label{fig:continuum-inference}
\end{figure}

Before applying PCA cleaning, we first test whether the broadband continuum
component can be recovered directly from the wavelength cross-power matrix.
Figure~\ref{fig:continuum-inference} shows the posterior constraints on the
two continuum-sector parameters for two instrumental-noise levels,
$\sigma_{\rm inst}=10^4$ and $10^3\,{\rm Jy\,sr^{-1}}$. In both cases the
posterior peaks are consistent with the fiducial input values,
$A_{\rm cont}=1.0$ and $\alpha_{\rm cont}=0.5$, with the fiducial point lying
within the central posterior contours. This demonstrates that, before
foreground cleaning, the wavelength cross-power matrix contains enough
information to recover the continuum model self-consistently. The constraints
tighten as the instrumental noise is reduced. For
$\sigma_{\rm inst}=10^4\,{\rm Jy\,sr^{-1}}$, the marginalized $1\sigma$
uncertainties are approximately
$\sigma(A_{\rm cont})\simeq 0.0015$ and
$\sigma(\alpha_{\rm cont})\simeq 0.0038$. For
$\sigma_{\rm inst}=10^3\,{\rm Jy\,sr^{-1}}$, these shrink to
$\sigma(A_{\rm cont})\simeq 0.0007$ and
$\sigma(\alpha_{\rm cont})\simeq 0.0016$. Thus the continuum amplitude and
broad spectral-tilt parameter are recovered at the sub-percent level in this
pre-cleaning continuum-only test.

The remaining degeneracy is between $A_{\rm cont}$ and
$\alpha_{\rm cont}$. This is expected because both parameters affect the
broadband continuum contribution to the wavelength cross-power matrix:
$A_{\rm cont}$ rescales the overall continuum power, while
$\alpha_{\rm cont}$ changes its smooth spectral tilt across wavelength. The
degeneracy is not exact because the tilt changes the relative power between
different wavelength-channel pairs, whereas the amplitude rescales the
continuum contribution more uniformly. This spectral-shape information allows
the continuum amplitude and broad color correction to be constrained
simultaneously before PCA cleaning.

\begin{figure}[t]
    \centering
    \includegraphics[width=\linewidth]{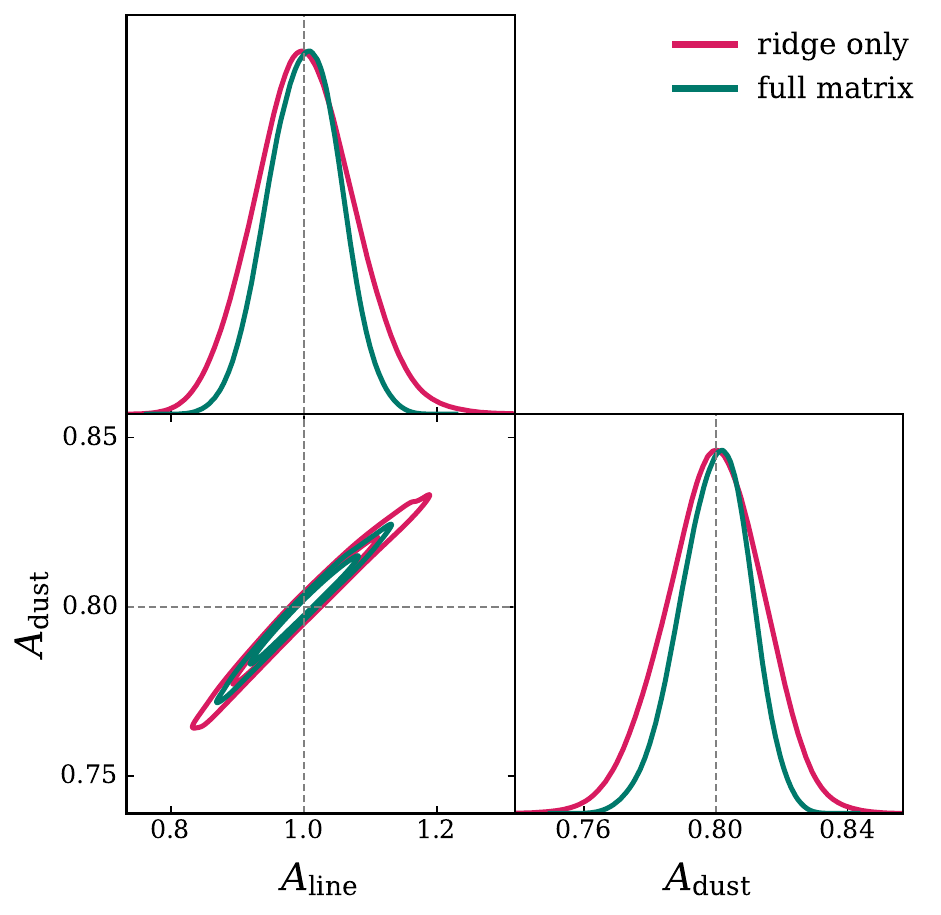}
    \vspace{0.35cm}
    \includegraphics[width=\linewidth]{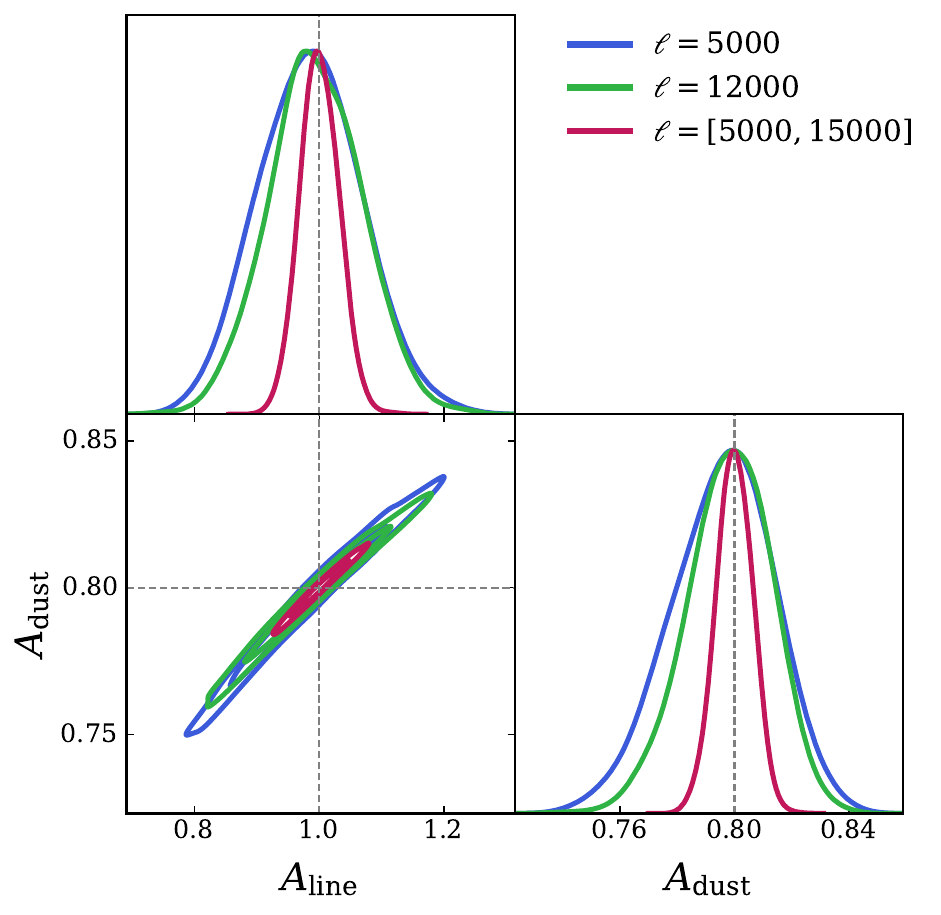}
    \caption{
    Post-cleaning inference of the LIM line-sector parameters using the forward PCA-cleaning model.
    Top: constraints on the line amplitude, $A_{\rm line}$, and dust parameter, $A_{\rm dust}$, comparing the ridge-only and full-matrix analyses at fixed multipole.
    Bottom: comparison of constraints obtained from different multipole choices, including single-$\ell$ and multi-$\ell$ analyses.
    The agreement between the contours and the fiducial values shows that the forward-cleaned model can recover the injected line signal, while the multi-$\ell$ case illustrates the additional information gained by combining angular scales.
    }
    \label{fig:lim-inference}
\end{figure}

After validating the continuum sector, we apply the forward
PCA-cleaning model to the cleaned maps and infer the line-sector
parameters from the cleaned cross-power matrix.  Figure~\ref{fig:lim-inference}
summarizes this recovery test for the line amplitude, \(A_{\rm line}\),
and the effective dust parameter, \(A_{\rm dust}\).  The numerical
posterior summaries are given in Table~\ref{tab:prepost_inference}.
The top panel compares two choices of data vector at fixed multipole:
the same-redshift ridge elements and the full selected
wavelength--wavelength matrix.  The ridge-only fit uses the part of the
matrix most directly associated with multi-line correlations at a common
redshift, while the full-matrix fit also includes additional
cross-channel information.  The two posteriors are consistent with one another and are centered near the injected model, demonstrating that the forward-cleaned templates successfully recover the line signal after removing PCA modes. The full-matrix posterior is narrower, as it should be, since it relies on more matrix elements, and therefore more information regarding the relative wavelength structure of the cleaned signal. This comparison is crucial as it reveals whether the outcome arises solely from the readily apparent ``ridge'' pixels or whether the larger cleaned matrix carries consistent line information.

The contours further indicate a positive correlation between \(A_{\rm line}\) and \(A_{\rm dust}\). Increasing the line amplitude raises the overall line power, while
changing the dust parameter changes the relative strength of the
dust-attenuated line template. Over the range of wavelength and multipole scales used here, these influences partly offset each other, resulting in inclined posteriors. Despite this degeneracy, the posteriors still remain localized, confirming that the multi-line matrix contains sufficient spectral information to distinguish between an overall line amplitude and an effective dust attenuation parameter. 

The lower panel examines the effect of increasing angular scale information on the posteriors. 
The single-\(\ell\) posteriors indicate information content at each individual angular scale, whereas the multi-\(\ell\) posterior integrates the cleaned cross-power matrices across a span of multipoles. 
The multi-\(\ell\) constraint is tighter, particularly in \(A_{\rm line}\), which signifies that the analysis benefits not only from the ridge's wavelength structure at a given scale, but also from the angular scale dependence of the cleaned angular power spectra. This is the primary conclusion from the multi-\(\ell\) comparison: combining different multipoles transforms it from a single-slice consistency test into a more informative LIM likelihood. Altogether, the two panels illustrate that the forward PCA-cleaning framework accurately reproduces the injected line-sector parameters post-continuum removal. The ridge-only analysis offers a conservative verification based on the most physically clear same-redshift line correlations, while the full-matrix and multi-\(\ell\) analyses highlight the advantages gained by incorporating additional wavelength and angular-scale information.

\begin{figure}[t]
    \centering
    \includegraphics[width=\linewidth]{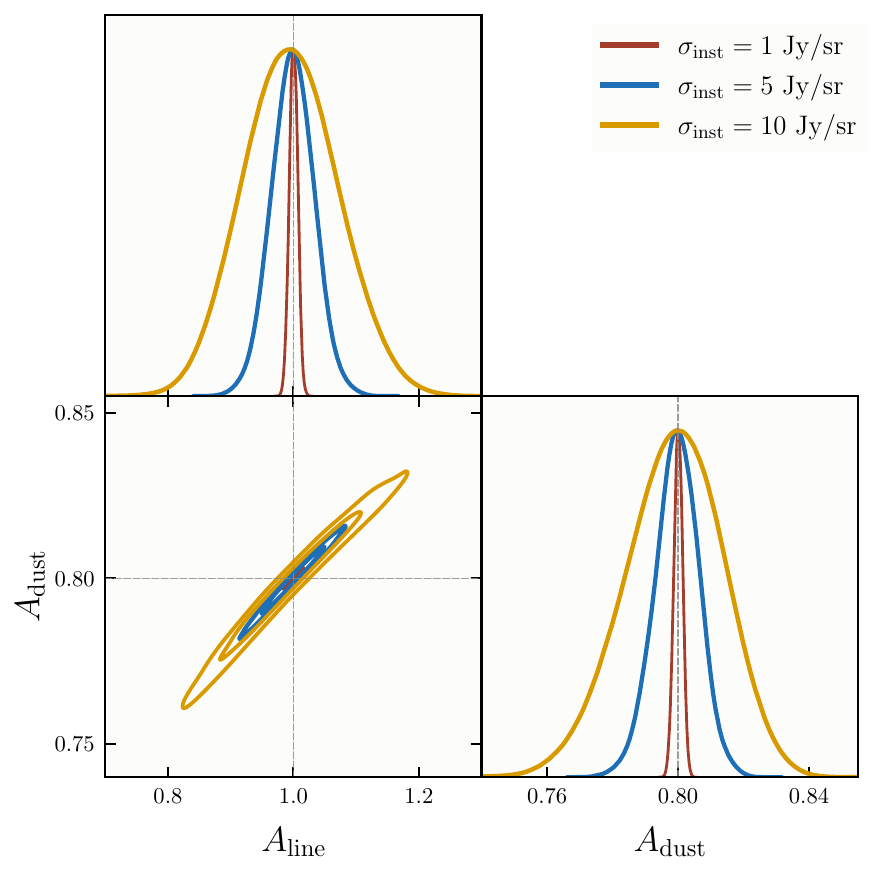}
    \caption{
    Dependence of the post-cleaning LIM parameter constraints on the instrumental noise level. 
    The contours show the inferred line amplitude, $A_{\rm line}$, and dust parameter, $A_{\rm dust}$, after forward PCA cleaning for different choices of the instrumental noise amplitude. As the instrumental noise increases, the posterior broadens, while the degeneracy direction remains similar. This confirms that the inference responds sensibly to the signal-to-noise level of the cleaned maps.
    }
    \label{fig:lim-inference-inst}
\end{figure}

In Figure~\ref{fig:lim-inference-inst}, we show how the inferred line-sector parameters respond to changes in the assumed instrumental noise level. In this test, the instrumental noise is scaled relative to the fiducial value while keeping the same forward PCA-cleaning and inference pipeline. As expected, lower instrumental noise results in smaller constraints for both $A_{\rm line}$ and $A_{\rm dust}$, whereas larger instrumental noise broadens the posterior contours. The figure also shows that the degeneracy direction between $A_{\rm line}$ and $A_{\rm dust}$ remains similar as the noise level is changed. This indicates that the same physical tradeoff is present in each case: the line amplitude controls the overall normalization of the line signal, while the dust parameter changes the relative attenuation encoded in the line template. The main effect of increasing instrumental noise is therefore to reduce the constraining power, not to qualitatively change the structure of the posterior.

\section{Discussion}
\label{sec:discussion}

In the above sections we showed that the shape of the resulting wavelength-wavelength grid that encodes power spectrum values at various scales depends on three coupled effects: intrinsic multi-line emission from the galaxy population, dust attenuation within these galaxies, and the much brighter continuum emission that acts as a foreground for MLIM observations. In this section, we summarize the physical meaning of these effects and consider what can be robustly estimated from the cleaned LIM observables.

\subsection{Information about Dust in the Cross-Channel LIM}

The results above show that dust attenuation leaves a measurable imprint
on MLIM observables, but that imprint is not equivalent to a
single global rescaling of the line power. Depending on the redshift, optical depth, inclination, and line wavelength, the lines are attenuated to different extents by dust in individual galaxies, which changes both the absolute normalization of the lines and the relative intensities of lines at the same redshift. This is the reason why the cross spectra $C_\ell(\lambda_i,\lambda_j)$ contain more direct information about the dust attenuation than the dimensionless correlation matrix $r_\ell^{ij}$: the latter is important to establish the geometrical ridge structures, but its normalization cancels out the amplitude information sensitive to attenuation.

Probing the structure of the multi-line cross-channel spectra is therefore essential for studying dust attenuation.  A single line mainly
constrains a combination of luminosity normalization, bias, and
attenuation.  By contrast, the joint pattern of H$\alpha$, H$\beta$,
[O~{\sc iii}], and [O~{\sc ii}] correlations compares transitions with
different rest wavelengths and different dependencies on the underlying
galaxy population.  The dust parameter used in the inference should
therefore be interpreted as an effective nebular attenuation parameter
for the LIM data vector, not as the dust optical depth of an individual
galaxy.  Its constraining power comes from the relative redistribution of
power among the line ridges and from the scale dependence of the cleaned
cross-spectra.

\subsection{Continuum Dominance and the Role of PCA Cleaning}

A fundamental issue is that the spectral continuum is much more luminous than the line emission in the raw wavelength maps. This results in broad, smooth correlations across wavelength due to the continuum emission, and narrow ridges along same-redshift lines due to the line emission. This difference in spectral structure enables the blind cleaning procedure. However, PCA cleaning of the continuum causes the loss of the signal. Since the spectral modes removed are extracted from the total observed data cube, they are comprised of continuum emission and some amount of line emission. Therefore, the transfer function is not a simple cosmetic correction but rather a part of the forward model.

The PCA diagnostics show there is a stable regime of acceptable cleaning parameters instead of a single optimum. If too few modes are removed continuum residuals will remain, while too many removed modes will suppress line emission itself. At the target multipole $\ell\simeq11571$, PCA15--PCA25 defines the useful cleaning range, with the transfer-quality metric maximized at PCA20. PCA25 provides a more aggressive robustness case with stronger cleaning. The two-dimensional scan in $(\ell,N_{\rm PCA})$ confirms this behavior at intermediate multipoles, meaning the scale is not an isolated numerical coincidence.
Since the transfer function $T_i(\ell)$ is channel-dependent this fact becomes important for quantitative analysis. While the median transfer statistics in the PCA diagnostic plot are useful for comparing PCA choices of $N_{\rm PCA}$, the proper likelihood must take into account the wavelength-dependent response of the cleaned maps. Particularly, the transfer should be negligible for small regions of wavelength, which cannot be corrected by multiplicative gain; in such survey analysis they should be downweighted or even discarded.

\subsection{Shot Noise, Clustering, and the Choice of Data Vector}

The decomposition into clustering and shot terms clarifies which parts of
the wavelength--wavelength matrix are of physical interest. The shot term is associated with emission from a single object appearing in many channels at the same redshift, and is localized primarily along the ridges on the same redshift axis. The clustering term arises from correlations between different objects across the large-scale density field and can exist throughout the matrix. This also explains why the shot noise maps have ridge-like structures, while the clustering maps have fluctuations across all regions.

This motivates the choice to compare the results using only ridge elements against those obtained from the full matrix. The data vector comprised only of ridge elements isolates the most direct multi-line correlations and has a clear physical interpretation. However the full matrix also encodes information from the off-ridge elements, continuum residual structure, and the overall pattern associated with the cleaned large-scale field. We demonstrated with the forward-PCA mock observations that the fit using the full matrix results in posteriors that are more tightly constrained compared to the ridge-only fit and are consistent with the input parameters. 
This shows that there is potential information content in the off-ridge matrix elements as long as one models the continuum residual structure and cleaning transfer accurately.

\subsection{Implications for Parameter Inference}

The inference tests show what the forward model is doing in a clean test of the situation where we have the truth values. In the data prior to cleaning, the raw matrix is overwhelmingly dominated by continuum power, and the continuum-sector parameters may be estimated accurately by the mock recovery. After PCA cleaning, the line-sector parameters can be reconstructed from the cleaned matrix when the line transfer and residual continuum contamination are part of the model. As expected, we find a significant posterior correlation between $A_{\rm line}$ and $A_{\rm dust}$ since increased true line amplitude and increased transmission both increase the perceived line power, though by differently affecting the perceived multi-line template ratio.

We see the advantage of using the multipoles, compared to single-$\ell$ information, in the constraint provided: The LIM information is encoded not only in wavelength space, but also in the angular-scale dependence of the ridge structure, which the multi-$\ell$ case constrains as a shape dependence in the angular power spectrum shape. This represents a significant step toward a real analysis, where the real data vector should be a combination of all useful multipole and wavelength-pairs with a covariance determined from mocks or an approved analytical approximation.

\begin{figure}[htbp]
    \centering
    \includegraphics[width=0.47\textwidth]{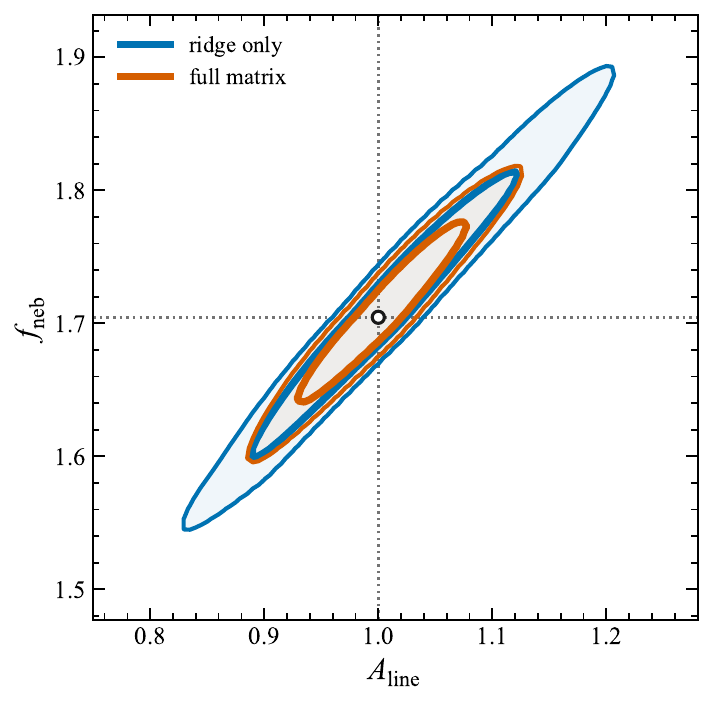}
    \caption{
    Recovery of the nebular dust parameter after PCA cleaning.  The contours show
    the joint posterior of the line-amplitude parameter $A_{\rm line}$ and the
    effective dust parameter $f_{\rm neb}$, using the forward-cleaned PCA20
    templates at $\ell\simeq11571$.  Blue contours use only the same-redshift
    line-ridge elements of the wavelength--wavelength matrix, while orange
    contours use the full selected matrix.  The dotted lines mark the injected
    truth values.  The elongated diagonal direction shows the expected
    amplitude--dust degeneracy: a larger dust attenuation suppresses the line
    signal and can be partly compensated by a larger line normalization.  The
    full-matrix fit narrows the allowed region relative to the ridge-only fit,
    indicating that off-ridge wavelength pairs add useful information after
    continuum cleaning.
    }
    \label{fig:physical-fdust-recovery}
\end{figure}

We then rewrite the dust parameter for this final test in terms of the nebular-to-stellar attenuation factor $f_{\rm neb}$, i.e. $A_{V,g}^{\rm neb} = f_{\rm neb}A_{V,g}^{\rm star}$. As a companion to the template-amplitude test with $A_{\rm dust}=0.8$, we inject an off-fiducial value $f_{\rm neb} = 0.75\,f_{\rm neb,0} \approx 1.7$, where $f_{\rm neb,0}=1/0.44 = 2.27$, and we treat the template-interpolation as being linear in $f_{\rm neb}$ over this limited range. Figure~\ref{fig:physical-fdust-recovery} displays the most ``clean'' implementation of the physical dust-recovery test. Here, we hold fixed the continuum sector and PCA cleaning response and determine whether the cleaned MLIM vector can recover the injected nebular dust parameter. In this case, we display the posteriors in the parameter space of overall line normalization, $A_{\rm line}$, and the effective dust parameter, $f_{\rm neb}$.

The primary feature of Figure~\ref{fig:physical-fdust-recovery} is the clear diagonal degeneracy between these two parameters. Physically, this degeneracy makes sense as increasing $f_{\rm neb}$ will suppress the nebular line emission, while increasing $A_{\rm line}$ will boost the effective overall line amplitude. As a consequence, the data constrain a combination of the strength of the line and its attenuation rather than each parameter independently. We show that the injected ``truth'' value lies within the recovered contours of this degeneracy space and is thus recovered without the information about dust being removed by the forward PCA-cleaning model in this idealized test.

Comparing the ridge-only fit to the full-matrix fit gives insight into the information content of the different portions of the data. The ridge-only fit makes use of the same-redshift pair structure, which already encodes considerable information on the nebular dust as the relative strengths of these features change with attenuation. The full matrix fit leads to a less degenerate region, illustrating that the use of multiple wavelength pairs beyond those within each same-redshift group can indeed help reduce the $A_{\rm line}$-$f_{\rm neb}$ degeneracy. The implication is that after accounting for continuum fitting and transfer, the LIM matrix captures more information about nebular dust attenuation than merely a detection of the total line amplitude.

This example inference serves as an illustration of the scientific use case for the pipeline: not just cleaning a LIM cube, but retaining enough multi-line detail that real physical parameters can be inferred from the cleaned cross-channel spectra. In our idealized case, the recovered posterior contours for $A_{\rm line}$ and $f_{\rm neb}$ show how attenuation by dust does leave a discernible signature in the relative amplitudes of the multiple line features that will ultimately be used. The presence of a degenerate relationship with the line normalization makes physical sense: it represents the degeneracy between decreasing the flux due to dust and simultaneously increasing the line amplitude. The fact that the full matrix fit tightens this degeneracy provides further evidence that MLIM encodes information about dust beyond what could be recovered from a single-line-power detection alone, and that non-ridge-pair wavelength measurements contribute to the determination of both normalization and dust properties.

\subsection{Limitations and Outlook}

This work serves as a first end-to-end validation of the modeling and inference framework for a LIM survey, not as a survey forecast. There are many simplified assumptions. Our mock lightcone spans a finite area, and the nebular emission lines are modeled with simplified empirical scaling laws. 
Furthermore, the dust model employed in this work is also idealized. 

In future work, we plan to implement more sophisticated modeling of the nebular emission based on the ISM properties in our model galaxies, supplemented by subgrid recipes for ionized H\,\textsc{ii} regions and by photoionization models. More complex modeling of the continuum and nebular dust will also be implemented, with this modeling calibrated against radiative transfer predictions. Our instrumental-noise tests are helpful in understanding the scaling behavior, but a full survey forecast should incorporate the survey mask, beam, calibration uncertainties, zodiacal and Galactic backgrounds, realistic variations in survey depth and a covariance matrix built from many independent realizations. Our main methodological message from this study is that the continuum cleaning procedure and parameter inference cannot be considered independently of one another. 
Our cleaned map is not the ultimate observable; it is a transformed product with a non-trivial dependence on the underlying line field. 
Subsequent work should address the full covariance of the cleaned cross spectra, confirm the channel-dependent transfer function with separate simulations and assess if our analysis framework can simultaneously recover both line and dust parameters in more survey-realistic mocks. In addition to that, we plan to apply the PCA cleaning and parameter inference directly to real SPHEREx data when it is made public.

\section{Conclusions}
\label{sec:conclusion}

We build a forward modeling framework for dust, stellar continuum
removal, and MLIM on a SPHEREx-like mock
survey. Starting from a SAM lightcone, we make line and continuum maps
for H$\alpha$, H$\beta$, [O~{\sc iii}], and [O~{\sc ii}], measure
cross-channel angular power spectra, perform blind PCA continuum
removal, and test parameter recovery with a forward-cleaned
likelihood. We summarize our main results below:

\begin{enumerate}

\item The line-only cross-channel matrix exhibits clear geometric
ridge structure, which occurs when two different rest-frame lines
enter two observed channels at the same redshift. This means the
normalized correlation matrix can be used as a diagnostic for the
redshift geometry of the multi-line signal.

\item Dust attenuation alters both the dimensional line power spectra
and the relative ridge amplitudes. A simple global scaling does not
work because dust attenuation is not monolithic; it varies with
wavelength, galaxy properties, and which galaxies contribute to a
particular line. Thus, the cross-channel absolute power spectra and
the relative amplitudes of the ridges encode information about dust
for inference.

\item Shot noise and clustering have differing wavelength-space
structures: shot noise is localized on same-redshift ridges because
both channels require the same galaxy, whereas clustering represents
density-field correlations over large spatial scales and is more
diffuse across the channel--channel matrix.

\item Raw maps are dominated by the stellar continuum, but the
underlying line-ridge structure is discernible after PCA
cleaning. PCA cleaning is a compromise between removing continuum and
preserving line signal; for the targeted multipole $\ell\simeq11571$,
PCA15--PCA25 is a reasonable range, and PCA20, which also maximizes
the scan's transfer quality, is our fiducial option. As a more
conservative choice for exploring robustness, PCA25 is also
presented.

\item Line transfer functions must be accounted for throughout the
analysis because PCA cleaning degrades the line signal to varying
degrees depending on wavelength. Median transfer quantities are
helpful for summarizing the effects of the cleaning, but accurate
inference requires the channel-dependent transfer response. Channels
with very low transfer should be down-weighted.

\item We test mock recovery of parameters for both the continuum and
line sectors in controlled setups. The continuum-sector parameters
are well recovered prior to cleaning. After cleaning, and by modeling
the forward response as the cleaned line response plus residual
continuum contamination, we are able to recover the line-sector
parameters accurately. The full-matrix fit produces tighter
constraints than the ridge-only fit, and further improvements are
seen with additional multipoles.

\end{enumerate}

These results indicate that nebular dust attenuation can be probed
statistically with MLIM in a survey like SPHEREx, even
though the stellar continuum is significantly stronger. This is a
first step toward leveraging LIM as a probe of dusty galaxies rather
than only a means of detecting their emission. This work provides a
basis for future survey-level analyses using realistic covariance
matrices and observational simulations.

\begin{acknowledgments}
AR acknowledges support from NASA under award number 80NSSC18K1014939. ARP was supported by NASA under award numbers 80NSSC18K1014, NNH17ZDA001N, and
80NSSC22K0666, and by the NSF under award number 2108411. The Flatiron Institute is supported by the Simons Foundation.
\end{acknowledgments}

\bibliographystyle{aasjournal}
\bibliography{references}

\end{document}